\documentclass[12pt]{article}
\usepackage{amssymb,amsmath} 
\usepackage{graphicx}
\graphicspath{ {figures/} }
\usepackage{epsfig} 
\usepackage{latexsym}
\usepackage{psfrag}   
\usepackage{subfigure}  
\usepackage{booktabs}  
\usepackage{braket}
\usepackage{mathtools}
\usepackage{textcomp}
\usepackage{ifthen}
\usepackage{comment}
\usepackage{protosem}
\usepackage{wasysym}

\usepackage{mathrsfs}
\usepackage[inner=2.9cm,outer=2cm]{geometry}
\usepackage[toc]{appendix}
\usepackage{color,soul} 
\usepackage{datetime}
\usepackage[
      colorlinks=true,
      linkcolor=darkblue,  
      urlcolor=black,    
      filecolor=blue,     
      citecolor=darkgreen, 
      linktocpage=true,
      pdfstartview=FitV,
      bookmarksopen=true    
      ]{hyperref}

\usepackage{lipsum}                     
\usepackage{xargs}                      
\usepackage[pdftex,dvipsnames]{xcolor}
\definecolor{darkblue}{rgb}{0.2, 0, 0.8}
\definecolor{darkgreen}{rgb}{0.2, 0.71, 0}

\newcommand{\GN}{G_{\text{N}}}
\newcommand{\ie}{i.e.}
\newcommand{\ra}{\rangle}
\newcommand{\la}{\langle}
\newcommand{\tr}{\text{tr}}
\numberwithin{equation}{section}

\newenvironment{changemargin}[2]{%
\begin{list}{}{%
\setlength{\topsep}{0pt}%
\setlength{\leftmargin}{#1}%
\setlength{\rightmargin}{#2}%
\setlength{\listparindent}{\parindent}%
\setlength{\itemindent}{\parindent}%
\setlength{\parsep}{\parskip}%
}%
\item[]}{\end{list}}

\begin{document}  


\begin{titlepage}

\vspace*{1.2cm}

\begin{center}
{\Large \bf Bekenstein bound in the bulk and AdS/CFT 
} \\

\vspace*{1.2cm}
 Irfan Ilgin\footnote{i.ilgin@uva.nl} \\
\medskip\vspace{0.5cm}
Institute for Theoretical Physics \\
University of Amsterdam\\
Science Park 904 \\
 1090 GL Amsterdam \\
 The Netherlands

\bigskip

\end{center}

\vspace*{0.1cm}

\begin{abstract}  
\begin{changemargin}{-0.95cm}{-0.95cm}
\noindent  
In this paper, we identify the change in the boundary full modular hamiltonians with the bulk observables for spherically symmetric excitations. The identification is demonstrated for perturbative as well as non perturbative excitations.  We introduce the notion of the \textit{sphere of ignorance}, that describes the bulk region that can not be probed by boundary regions below a certain size. It is argued that the vacuum subtracted entropy in the bulk associated with the sphere of ignorance is bounded by the difference of the change of entanglement entropies for complementary regions in the boundary for spherically symmetric state. Bekenstein bound for the sphere of ignorance reflects itself in the boundary theory as the positivity and monotonicity of the relative entropy of the complementary boundary balls. We compare the proposed bound with Araki-Lieb bound and identify the non-trivial domains where Bekenstein limit sets the lower bound. Moreover, we clarify throughout the paper fundamental differences between pure state and thermal excitations from an information theoretic point.
\end{changemargin}
\end{abstract} 

\end{titlepage}

\setcounter{tocdepth}{2}
{\small
\setlength\parskip{-0.5mm} 
\tableofcontents
}


\newpage
\section{\label{sec:level1}Introduction}

Bekenstein bound \cite{Bekenstein:1980jp} is the universal upper bound on the entropy $S$ or information that  can be contained in a physical system or object with given size and total energy. If $R$ is the radius of a sphere that encloses a given system, while $E$ is its total energy including any rest masses, then its entropy is bounded by,
\begin{align}
    S \le \lambda R E
\end{align}
where $\lambda$ is a numerical constant of order one. Although the derivation of the bound uses \textit{generalized second law} around black holes \cite{Bekenstein:1974ax}, the bound seems to be independent of the gravitational physics. This fact manifests itself in $\GN$ independence of the bound. Moreover the size of the box $R$ is the geodesic distance in flat space. These observations indicate that the Bekenstein bound is valid in flat space hence can be derived via information theoretic inequalities in QFTs. In \cite{Casini:2008cr} the bound is derived by employing positivity of relative entropy for certain class of excited states with respect to vacuum. The derivation exploits the local expressions of modular Hamiltonians of certain spatial sections of vacuum density matrices. On the other hand, the Bekenstein bound manifests itself also on the systems having strong self gravitation. It is a well known fact that the bound is saturated for the Schwarzschild black hole. In other words the Bekenstein bound is saturated when  Schwarzschild energy is put into a box of Schwarzschild radius. This is a strong indication that the bound preserves its validity beyond the weak self gravitating systems and hence should have a formulation for systems having back-reactions. One difficulty that is encountered when the system has back-reaction on the spacetime, we don't have at hand a natural definition of the size of the box. For example, in the case of a Schwarzschild black hole the radius of the box is not geodesic distance but the radial coordinate corresponding the black hole horizon. Because of the above observations, natural questions arise. How to define the radius of the box $R$ in the presence of back reaction? What is the energy of the system $E$ for a strong self gravitating system or in a setup that allows backreaction? Let us  emphasize that for a black hole, size of the box $R$ is the coordinate radius and energy of the system is the ADM mass of the solution which includes the binding energy of the entire solution. In a way, size of the box is determined with respect to the vacuum solution. Vacuum solution provides a reference grid where excited system can be compared to. Our main goal is to identify the boundary information theoretic observables corresponding the Bekenstein bound in the bulk including systems having non perturbative backreactions on the spacetime metric, such as black holes. We will study the problem using AdS/CFT \cite{Maldacena:1997re, Witten:1998qj, Aharony:1999ti} and find the corresponding information theoretic inequality on the CFT that describes the bound in the bulk. We will give the formulation of Bekenstein bound in the bulk for certain class of excitations on asymptotically AdS$_{d+1}$ through the information inequalities in the dual CFT.

Before giving the formulation of the Bekenstein bound in the bulk via the underlying theory, we will clarify issues regarding the first law of entanglement entropy on a simple exercise involving conical defects. This will be a simple demonstration on how quantities involved in the first law type relations in CFT are identified with the quantities in spacetime. In this identification local expressions for modular hamiltonians \cite{Bisognano:1976za} and Ryu-Takayanagi formula \cite{Ryu:2006bv, Ryu:2006ef, Lewkowycz:2013nqa} play the central role. The so called first law of entanglement due to presence of conical defects provides us a puzzle which we address in the third section. The solution of this puzzle will also clarify the differences between pure state and thermal (in general mixed state) perturbations when first law of entanglement is considered. The complete knowledge of the pure state puts strong constraints on the expectation value of local stress energy on the dual CFT. Such constraints do not exists for mixed state perturbations as we will demonstrate.

In the forth section we have exercised the initial conical defect setup using the fundamental expression of covariant phase space formulation \cite{Iyer:1994ys} of the first law of black hole thermodynamics. The formulation is valid for perturbative excitations\footnotemark[1]
 \footnotetext[1]{Along the paper, perturbation is considered for two cases. It is either indicating a perturbation on the underlying state, $\ie$ deformation of the state into a nearby one in the Hilbert space or the metric field is perturbed by a $\delta g$. As it will be shown these two cases do not have to match at every order of perturbation.}  and conical defect solutions can be studied in this regime. In this formulation, origin of the differences between the relative entropies of the complementary regions will be clear. Due to the pure state nature of the conical defect perturbation, one can reduce the differences between relative entropies of the complementary spherical regions ($\{A, \bar{A}\}$) on the boundary into differences of modular hamiltonians. Moreover the differences in modular energies,  $\Delta H_{\bar{A}}-\Delta H_{A}$ can be identified with the bulk modular hamiltonian in the perturbative regime. Using the boundary expression of differences between modular energies of the complementary spatial sections we extend the notion of bulk modular energy of a spherical region around the `origin' to non perturbative excitations. The result of the calculation of the vacuum subtracted full modular hamiltonian for excited states are very elegant and does not depend on the dimension of the spacetime. Here we introduce the notion of \textit{sphere of ignorance} which is the bulk region that is not accessible from any boundary interval having size below a certain length scale. In other words an observer having access to the region $A$ in the underlying theory can not decode anything in the bulk beyond the bulk sphere $< R_{\text{scale}}$. The essential relation we derive is that the modular energy contribution of the bulk excitation depends linearly to the scale of the system that contains it. This is closely related with UV-IR correspondence in AdS/CFT, and will be explained in more detail throughout the paper. The full modular hamiltonian\textemdash which we also named as \textit{bulk modular energy} inside the sphere of ignorance\textemdash for spherical excitation is given by

\begin{align}\label{mainresult_1}
\Delta H_{\bar{A}-A} = 2\pi R_{\text{scale}}\, \Delta M_{\text{ADM}} .
\end{align}
The definition of full modular hamiltonian is $\hat{H}_{\bar{A}-A}\equiv  \hat{H}_{\bar{A}} - \hat{H}_{A}$ and $R_{\text{scale}}$ is the radial position of the tip of the minimal surface that is homologous to region $A$ on the boundary. 

\begin{figure}[t!]
    \centering
    \includegraphics[width=0.75\textwidth]{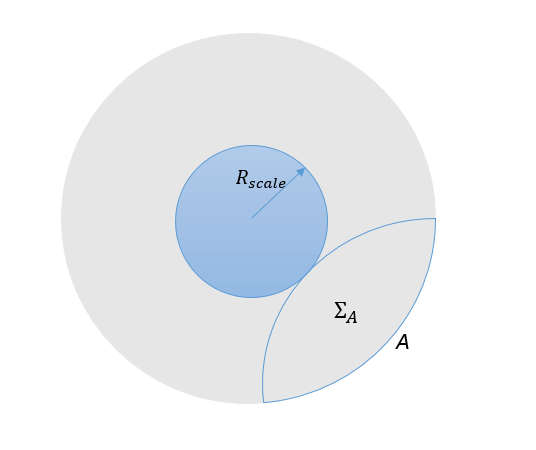}
    \caption{Blue sphere denotes the region that observers who have access to the boundary region of size($A$) are blind to. Radius of the sphere, $R_{scale}$, is the deepest point that can be decoded by accessing $A$. In a spherically symmetric state, sub-states having support on the region of same size, have the same change in information content, hence the only region that can not be decoded by the observer $A$ is the blue sphere which is referred as the sphere of ignorance along the paper.}
    \label{bekenstein_bound}
\end{figure}
$\Delta M_{\text{ADM}}$ is the vacuum subtracted ADM mass. The expression is very elegant as it is valid in any dimension. The above expression is not only valid perturbatively but holds also for excited states that can not be expressed as infinitesimal deformations of the vacuum. In the non perturbative case, $R$ denotes the radial position of the point of ignorance \textemdash tip of the geodesic \textemdash with respect to the global vacuum. 

A similar expression is used in \cite{Verlinde:2016toy} as a change of area  in the weak field limit for certain identification of the manifolds and interpreted as the amount of entanglement entropy reduced by bulk excitation from its surrounding. In our interpretation, it quantifies the modular energy contribution of the bulk excitation to the entanglement wedge that contains it. The presence of bulk excitation in the entanglement wedge is the source of the differences between modular energies of complementary regions in the underlying theory.

For the perturbations around AdS$_{d+1}$, one can relate the boundary modular energy with the bulk modular integral as explained above. However, this is only possible for pure state deformations around the vacuum as it is explicitly using the relation $\Delta S_{A} = \Delta S_{\bar{A}}$ for pure states.  Detailed explanation about this can be found in Section 4. 
On the other hand one can come up with a mathematically similar, yet physically quite different expression when the underlying state is thermal, or in general, a mixed state. In a mixed state, one can not constrain the change of entanglement entropy of the complementary regions via their equality. This allows a net change in the total energy of the system \cite{deBoer:2015kda} when the perturbation is of thermal nature, which we have detailed in section 3 starting from a conical defect exercise. Using the first law of entanglement entropy and freedom to increase total energy of the system in a mixed state perturbation we end up with the following expression for mixed states when the local bulk excitation is confined into the sphere of ignorance
\begin{align}\label{mainresult_2}
\delta S_{\bar{A}-A} = \delta S^{\text{bulk}}_{\mathbb{S}_{R}^{d-1}}= 2\pi R_{\text{scale}}\,\delta M_{\text{ADM}} 
\end{align}
where $\delta S^{\text{bulk}}_{\mathbb{S}_{R}^{d}}$ is the bulk entropy that corresponds to $1/N$ corrections to the entanglement entropy in the CFT \cite{Faulkner:2013ana}. We distinguish vacuum subtracted quantities when the state is perturbation around the vacuum $|\psi\rangle=|0\rangle+\epsilon |\phi\rangle$, by lowercase  $(\delta)$. This entropy resides inside the $d-1$-sphere of radius $R$. We find it interesting to compare the differences between pure states and thermal states excitations, since these differences are closely related with the volume law  \cite{Verlinde:2016toy} and ER-EPR conjectures \cite{Maldacena:2013xja}. 
It is crucial to realize the two main differences between these two equations. Firstly, while the eq. \eqref{mainresult_1} can not take place at the linear level of the perturbation on the underlying state, the later, \eqref{mainresult_2} can be derived only through using the first law of entanglement hence at the linear level. In other words in the first case the constraining equation is equality of change of entanglements for complementary states, in the second case the first law of entanglement entropy, $\ie$ equality of change of modular energy and entanglement entropy for each state. Therefore these two results have different physical principles behind them. Secondly the quantities involved in the relation, that is change of modular energy, $\Delta H_{\bar{A}-A}$ vs change in the entropy $\delta S_{\bar{A}-A}$. In the first case the bulk excitation does not possess any entropy in the form of entanglement with the purifying auxiliary system. The mixed state deformations of the vacuum at the linear level is particularly important since this is the only case where we can directly show that the entropy inside the sphere of ignorance is equal to the difference of change in the entanglement entropies of the underlying theory, $\delta S^{\text{bulk}} = \delta S_{\bar{A}-A}$. Equality holds when all the excitation is localized inside this sphere.

Finally and most importantly, we will extend the result for thermal states to non perturbative level where differences are defined with respect to vacuum. In this case one can not use the first law of entanglement entropy, which was used in \eqref{mainresult_2} to obtain the equality. In the generalization, we have expressed \eqref{mainresult_2} in terms of the differences of relative entropies of the complementary states. For a rotational invariant state imposing the positivity together with the monotonicity of relative one can come up with inequality version of equation \eqref{mainresult_2}. The differences of modular hamiltonians can be expressed in terms of the bulk quantities. This is the direct analog of the first equation \eqref{mainresult_1}. We have observed that difference of the change in the entanglement entropies of the complementary regions is bounded by the associated bulk modular energy, which was defined also through \eqref{mainresult_1}. The expression is interesting as it is strikingly reminiscent of the Bekenstein bound \cite{Bekenstein:1980jp} in the bulk. 
\begin{align}\label{mainresult_3}
\Delta S_{\bar{A}-A} \le 2\pi R_{\text{scale}}\,\Delta M_{\text{ADM}} 
\end{align}
The radius $R$ is interpreted on AdS$_{d+1}$. It is the radius of the ignorance sphere defined with respect to the boundary region $A$. We argue that for spherically symmetric state the entropy that resides inside the sphere of ignorance is bounded by the entropic differences of the complementary boundary balls $S^{\text{bulk}}_{\mathbb{S}_{R}^{d-1}}\leq\Delta S_{\bar{A}-A}$. We will read this expression backwards\textemdash in a sense\textemdash by asking what the information theoretic extensions of the observables of underlying theory to the bulk physics. It will be explained why the differences in entanglement entropies of the complements in the underlying state manifest itself as the Bekenstein bound in the bulk. Boundary perspective that is presented throughout the paper have some overlapping content presented in \cite{Blanco:2017akw} \cite{Herzog:2014fra} yet the bulk interpretation is completely novel according to our knowledge.

\newpage
\section{\label{sec:level1}A simple first law}  

We would like to start with a simple explicit example for the gravitational counterpart of the first law of entanglement. We will shed light onto interesting set of constraints for the first law entanglement entropy. Let us consider simple case of AdS$_3$ which will be sufficient for our purposes. 

First law of entanglement entropy holds for any state since the relative entropy vanishes at the linear level perturbatively. To see this, consider a reference state $\rho_0$ and an arbitrary state $\rho_1$. One can construct a family of interpolating density matrices
\begin{align}\label{relativentropylinearized}
\rho(\lambda)=(1-\lambda)\rho_0+\lambda\rho_1
\end{align}
where $\lambda$ can be positive or negative, yet the relative entropy $S(\rho(\lambda)|\rho_0)$ is positive for either sign of 
$\lambda$ by the positivity of relative entropy.  Hence first derivative of the relative entropy with respect to $\lambda$ vanishes. This is the first law of entanglement since relative entropy can always be expressed as,
\begin{align}
S(\rho(\lambda)|\rho_0)=\Delta \la \hat{H}_A \ra - \Delta S_A 
\end{align}
where $\Delta \la \hat{H}_A \ra = \text{tr}(\rho(\lambda) \hat{H}_A)-\text{tr}(\rho_0 \hat{H}_{A})$ and $\Delta S_A = -\text{tr}(\rho(\lambda)\log\rho(\lambda))-\text{tr}(\rho_0\log\rho_0)$. In the leading order of $\lambda$ relative entropy exactly vanishes, which is known as the first law of entanglement entropy,
\begin{align}
\delta S_{A}=\delta\langle \hat{H}_{A}\rangle
\end{align}
where we denoted the linear perturbation by lowercase delta $(\delta)$. However, there are only few cases where the local expression for $\hat{H}_A$ is known explicitly \cite{Cardy:2016fqc}. One such case is the ball shaped region  in the vacuum state of a CFT in any dimension. In this case modular hamiltonian has a local expression in terms of the local stress energy tensor of the CFT. 
\begin{align}
\hat{H}_{A} =2\pi \int_{A} \zeta^{\mu}\, \hat{T}_{\mu\nu}\, d\Sigma^{\nu}
\end{align}
where $\zeta^{\mu}$ is the conformal Killing vector that leaves the causal diamond of the ball shaped region invariant. This is an exact operator expression obtained by the conformal transformation from the half of the Minkowski space via employing the invariance of vacuum under  global conformal transformations. We will give a straightforward yet illuminating application of this local expression of the modular hamiltonian. 

Consider a 2d CFT on $\mathbb{R} \times \mathbb{S}^{1}$ with a classical gravitational dual. Suppose the vacuum is perturbed to a nearby pure state, $|\psi\ra = |0\ra + \epsilon |\phi\ra$, such that the perturbed state has uniform expectation value for energy density $\ie$ $\delta\la T_{00}\ra=\mu$. Let us consider a ball shaped region on the spatial slice $\ie$ a constant time slice of the CFT. For an interval on the boundary, explicit expression for the change of modular energy is given by,
\begin{align}\label{modularhamiltonianresultconicaldefect}
\delta\la\hat{H}_{A}\ra &= 2\pi \int_{A} \left(r\frac{\cos(\theta_0-\theta)-\cos(\alpha)}{\sin\alpha}\right)\, \la\hat{T}_{00}\ra\,r d\theta \nonumber\\
&=2r (1-\alpha\cot\alpha)\delta E_{\text{CFT}}
\end{align}
where $\theta_0$ is the center of the region in the boundary, $\alpha$ is the half of the total angular size of the boundary ball and $r$ is the radius of the $\mathbb{S}^{1}$. $\delta E_{\text{CFT}}$ is the change in the energy of the state, which is equal to the change in the total energy of spacetime, given by $2\pi r \mu$. The combination, $r \delta E_{\text{CFT}}$ is the dimensionless factor that we should expect to identify in the gravitational dual. 

The modular hamiltonian side of the calculation does not refer to the gravitational dual. We simply use the local expression for the modular hamiltonian on the CFT which is exact thanks to conformal symmetry. One can check the first law explicitly via holography, by identifying the geometric description of the state in the gravitational description. We have considered the perturbation to be a pure state with a uniform asymptotic energy density. The latter ensures the bulk solution to be a spherically symmetric one. One possible geometric description of the state is the conical defect. Conical defect geometries are 3d solutions of Einstein equation with a Dirac delta type source distribution. Although the expression for the change of modular energy is valid between any state and vacuum, the first law only holds for small perturbations around vacuum. Therefore we would like to look at change of entanglement entropy in the presence of a conical defect with small deficit angle which serves as the perturbation parameter. 

The metric of the conical defect is given by,
\begin{align}\label{conicaldefectmetric}
ds^2=-\left(\gamma^2+\frac{R^2}{L^2}\right) dT^2+\left(\gamma^2+\frac{R^2}{L^2}\right)^{-1}dR^2+R^2d\theta^2
\end{align}
where 0$<\gamma<$1 and related to the deficit angle as $\delta\theta=2\pi(1-\gamma)$. The minimal surface that is homologous to a boundary region $A$ measures the entanglement entropy of the subsystem that resides on $A$ on the dual CFT.
\begin{align}\label{conicaldefectentanglement}
 S(\alpha)=\frac{2L}{4\GN} \log\left(\frac{2L}{\gamma \epsilon}\sin(\gamma\alpha)\right)
\end{align}
$\epsilon$ corresponds to the UV cutoff in the underlying theory, and serves as a IR cutoff in the bulk that regularizes the infinite area of the boundary sphere in AdS. Although this expression is infinite in the limit that sends the cutoff to zero, the change of entanglement with respect to vacuum is finite once the cutoff is fixed. Looking at the change of entanglement entropy or any other quantity defined on different manifolds requires a comparison scheme. One such physical scheme is to keep number of degrees of freedom fixed since it is the characteristic of the theory describing these two states. From the spatial point of view, fixing degrees of freedom is to keep the ratio of the size of the system to the cutoff fixed. 
\begin{align}
\Delta S(\alpha) = \frac{2L}{4\GN} \log\left(\frac{\sin(\gamma\alpha)}{\gamma\sin(\alpha)}\right)
\end{align}
which in the small deficit limit, $\delta\theta/2\pi\ll 1$, perturbatively becomes,
\begin{align}\label{entanglemententropyresultconicaldefect}
\delta S(\alpha) = 2L(1-\alpha\cot\alpha)\delta M_{\text{ADM}}
\end{align}
since the change of ADM energy is $\delta M_{\text{ADM}} = \delta\theta/8\pi\GN$. Thus identification of $L \delta M_{\text{ADM}}\equiv r \delta E_{\text{CFT}}$ completes the demonstration of the first law. However, in the next section, we will have a closer look at this \textit{so-called} first law. As we will see, what seems like a first law is, in fact, not a first law from the information theoretic point of view. It does not concern the linear level of the parameter that connects density matrices of the underlying theory.

\section{\label{sec:level1}Puzzles about the first law}

In the previous section we have explicitly demonstrated, the first law of entanglement entropy through a gravitational calculation where excited state is considered to be a conical defect. We have matched the change of modular energy on the CFT to change of area of the minimal surface in the bulk due to appearance of a defect. Although it looks like we have computed different quantities in different theories, the RT conjecture maps them. A more careful look into what we have done will reveal a greater understanding. 

Initially, we consider AdS$_3$ solution with a boundary $\mathbb{S}^{1}\times \mathbb{R}$. Usually when one is restricted to a subspace on the boundary, there is no need to specify the change on the state having support on the complementary region to study the first law. Yet considering the subspace with its complement puts strong constraints on the $\delta\langle T_{\mu\nu}\rangle$. Remember that we considered a uniform perturbation on the vacuum, $\delta T_{00}(\theta)=\mu$ which is identified with conical defect solutions that has Dirac delta type sources in Einstein equation \cite{Deser:1983tn}. These solutions can be arbitrarily 
close to AdS$_3$ as it is possible to choose $\delta\theta/2\pi \ll 1$. Hence one can consider the bulk dual of the perturbation on the vacuum as a conical defect solution. Of course, this is just one particular solution with the given boundary energy density, one may come up with different semiclassical gravitational descriptions having same energy density. For example the perturbation could also be due to thermal fluctuations around vacuum in which case dual would be thermal AdS$_3$. We will study the thermal perturbations later on in this paper. First we will focus to the conical defect. It is crucial that, as explained under eq. (\ref{conicaldefectentanglement}), to be able match the change in modular energy to $\delta S(\alpha)=S(\alpha)_{\text{con.}}-S(\alpha)_{AdS_{3}}$ one needs to rescale the IR cut-off for the conical defect solution. This corresponds to keeping number of degrees of freedom fixed by fixing the proportions of UV cut-offs to the size of the systems where the underlying theory lives. To be explicit, if one considers the conical defect geometry as a angular cut, then one should rescale the UV cut-off on the boundary such that 2$\pi r /\epsilon_{UV}$ is fixed. This is one such beauty of AdS/CFT that it provides an unambiguous way to compare observables on nearby solutions.  

The question we would like to raise is if there exists a first law on each boundary interval as a result of conical defect type excitations. Before answering this question, let us consider a perturbation on the vacuum, which changes the state into another pure state that is infinitesimally close
\begin{align}\label{purestateperturbation}
|\psi \rangle=| 0 \rangle+\epsilon |\phi \rangle .
\end{align}
We didn't specify how the energy density of the perturbation is organized spatially. Suppose that, such a change in the state causes a localized linear perturbation $\delta \langle T_{\mu\nu}\rangle$.  If the perturbation is completely localized inside of a spherical region $A$ in the CFT  then change in the modular Hamiltonian of the complement $\bar{A}$ vanishes. The first law ensures that the change of entanglement entropy also vanishes, $\delta S_{\bar{A}}=0$. As you will realize, we end up with a contradiction. Because the perturbed state was also assumed to be a pure state. In that case
we would expect $\delta S_{A}=\delta S_{\bar{A}}$, yet by the first law of entanglement, $\delta S_{A}\neq0$. This would also violate the positivity of relative entropy, because $\delta \langle H_{A}\rangle=\delta S_A=\delta S_{\bar{A}}>0$, yet $\delta \langle H_{\bar{A}}\rangle=0$, which implies, $\delta \langle H_{\bar{A}}\rangle-\delta S_{\bar{A}}<0$. What is the resolution of this apparent paradox \cite{Casini:2011kv}? Indeed we have made a false assumption, by considering pure state perturbation localized only in a region at the linear level of the perturbation theory of the underlying state. One needs to put equal amount of energy to the complement in a pure state. This fact can be demonstrated simply through a field theory argument. Let us start by constructing the following operator,
\begin{align}
\hat{H}=\hat{H}_{A}-\hat{H}_{\bar{A}} .
\end{align}
This operator, known as \textit{full modular hamiltonian}, generates conformal transformations that keeps the spherical region fixed, hence annihilates the global vacuum state. The simplest way to understand why this operator annihilates the global vacuum is to consider the half space in QFT, where modular hamiltonian is the generator of rotation in the euclidean plane. Then the combination $H=H_{A}-H_{\bar{A}}$ generates a boost on the whole state, which leaves the vacuum invariant
\begin{align}
\hat{H} |0\rangle=(\hat{H}_{A}-\hat{H}_{\bar{A}})|0\rangle=0 .
\end{align}
Now if the state changes according to (\ref{purestateperturbation}), then
\begin{align}
\delta\langle H_{A} \rangle&=\epsilon\left(\la \phi | H_{A}| 0\ra+ \la 0 | H_{A} |\phi \ra\right)\nonumber\\
&=\epsilon\left(\la \phi | H_{\bar{A}}| 0\ra+ \la 0 | H_{\bar{A}} |\phi \ra\right)=\delta\langle H_{\bar{A}} \rangle
\end{align}
The above equality indicates that whenever one creates some some localized wave packets inside a region, to stay in a pure state, some energy density needs to be introduced outside. It is the purity of the state that enforces such constraint.

Let us further study what kind of constraints we have on the 
perturbation of the expectation value of stress energy tensor. Assume that constant time slice has the topology of $\mathbb{S}^{d-1}$ with radius
$r$. The modular Hamiltonian for $(d-2)$ dimensional spherical entangling surfaces surrounding a cap of the $S^{d-1}$ specified by the angle $\alpha$ is given by,
\begin{align}\label{d-dimensionalmodularhamiltonian}
H_{S_{A}^{d-2}}=2\pi\int^{\alpha}_{0}\,r^{d-1} \,d\Omega_{d-2} \sin^{d-2}\theta\,d\theta \left( r \frac{\cos\theta-\cos\alpha}{\sin\alpha}\right)  T_{00}(\vec{r})
\end{align}
This expression is the generalization of what we have used for CFT$_2$.  Now we can obtain $H_{\bar{A}}$ by sending the origin of the spherical cap, $\mathbb{S}^{d-1}$
to $\pi$ and $\alpha$ to  $\pi-\alpha$. Then the full modular Hamiltonian $H_{\bar{A}-A}\equiv H_{\mathbb{S}_{A}^{d-2}}-H_{\mathbb{S}_{\bar{A}}^{d-2}}$ becomes,
\begin{align}\label{fullmodularhamiltonian}
H_{\bar{A}-A}=2\pi\int^{\pi}_{0}r^{d-1} \,d\Omega_{d-2} \sin^{d-2}\theta \left( r \frac{\cos\theta-\cos\alpha}{\sin\alpha}\right) T_{00}(\vec{r})
\end{align}
For a pure state, by the first law of entanglement entropy, $\delta\langle H_{A} \rangle-\delta\langle H_{\bar{A}} \rangle=0$ at the linear level. Using this equality one can put constraints to the possible $\delta \langle T_{00}\rangle$ for pure state perturbations.  To see this explicitly, multiply $\delta\la H_{\bar{A}-A} \ra$ by
$\tan\alpha$ and take the derivative w.r.t to $\alpha$. The second term is eliminated and we have an example of such constraint which is
$\int d\Omega_{d-1} \cos\theta\, \delta \langle T_{00}\rangle=0$. Then using this one, we show easily, $\int d\Omega_{d-1}\, \delta\langle T_{00}\rangle=0$. These are not the only constraints because the first one is obtained by placing the origin of the cap on the $z$ axis. The full modular hamiltonian vanishes at the linear level independent of the choice of origin for boundary balls. It is true for all possible balls. Therefore one obtains the following set of 
constraints  \cite{deBoer:2015kda}
\begin{align}\label{constraints}
\int d\Omega_{d-1}\, \delta\langle T_{00}(\Omega)\rangle=0,\hspace{1cm}\int d\Omega_{d-1}\, \hat{\omega}\, \delta\langle T_{00}(\Omega)\rangle=0 .
\end{align} 
$d\Omega_{d-1}$ is the volume form on $\mathbb{S}^{d-1}$ and $\hat{\omega}$ is the unit vector parametrizing the points on  $\mathbb{S}^{d-1}$. Note that second constraint is generalization of $\int d\Omega_{d-1} \cos\theta\, \delta \langle T_{00}\rangle=0$ in which case one focuses on the $z$-component of $\hat{\omega}$. The first constraint resolves our initial confusion. One can not introduce local excitations on some regions without  balancing them with negative energy contributions. Another interesting distribution that is violating the second constraint is Dirac delta sources unless equal amount of positive and negative charges introduced at the same point, which is very constraining.  It would be interesting to see examples of distributions satisfying these constraints, yet we leave it for future studies. We are now ready to further puzzle ourselves with the first law in the presence of a conical defect.

If we now go back to our initial construction, where we consider the conical defect solution with a small deficit as the bulk dual of a homogeneous perturbation of the boundary theory around the vacuum. The change of area of the minimal surfaces due to conical defect was equal to $\delta \la H(\alpha)\ra$  for each boundary region. When $\delta \la T_{00}\ra$ is uniform then conical defect is at the center.  Their fusion yields us the generalization of the first(?) law to arbitrary surfaces which we present in a different paper. At this point we need to reexamine this first law in the light of the puzzles we had uncovered. 

First of all, as we have derived above, the first law of entanglement for any pure state perturbation puts some constraints on the total energy of the perturbed state such that it vanishes (\ref{constraints}). On the other hand, we expect the conical defect geometry to represent a pure state in the underlying theory, yet clearly its energy does not vanish. Therefore it violates the constraints above. Furthermore, assuming that for each boundary region, $A$, there is a first law, then we would expect that the state on $\bar{A}$, also satisfies the first law. However through the Wald formalism, it is clear that when 
perturbation is sourced by stress tensor then the first law is modified by contribution from stress energy of the perturbation, this will be further explained in the following sections. Hence one does not have a first law for the complement $\bar{A}$ whose entanglement wedge includes the defect.
In this case, one has $\delta H_{\bar{A}}-\delta S_{\bar{A}}\sim \delta$, where $\delta$ quantifies the deficit angle. Therefore either the perturbed state is not pure or this is not the first law of entanglement entropy. We know that the perturbed state is pure, since it is obtained from the vacuum by adding a localized non thermal mass. Let us be more careful about the order of perturbation. The first law takes place at the linear order in $\epsilon$, where fields perturbed according to $\phi\rightarrow \phi+\epsilon\,\phi^{(1)}+O(\epsilon^2)$. However the perturbation of the vacuum, by a classical mass distribution takes place in second order as $T_{\mu\nu}$ is
quadratic in fields. Indeed, what we have called as first law was taking place between the second order and first order. Let us look at it in more detail. 

From this point on, we will replace our notation for the variation with $\Delta$ to denote that difference is beyond the linear order. The perturbation 
of the solution by addition of classical matter, as explained above, takes place at the second order in perturbation theory. If the perturbed state is a pure state, $|\psi\ra=| 0\ra+\epsilon | \phi \ra+O(\epsilon^2)$, then the full modular Hamiltonian at this level does not vanish. Remember that the vanishing of that at the linear level yields us the constraints on the change of $\delta \la T_{00}\ra$, which does not exist beyond linear level.
\begin{align}
\la \psi | H | \psi \ra=\epsilon^2 \la \phi | H | \phi \ra+O(\epsilon^3)
\end{align}
Therefore we are not restricted to the constraints in (\ref{constraints}) at this level. Remember that the first law of entanglement entropy is derived
from the relative entropy at the linear level (\ref{relativentropylinearized}). Next to leading order, due to positivity of the relative entropy, one has,
\begin{align}
\Delta H_{A}-\Delta S_{A}\ge 0
\end{align}
The interesting point is that, when the perturbation is due to some localized mass distribution outside of the entanglement wedge $A$ one has $\Delta H_{A}=\Delta S_{A}$ at the leading order of perturbation by $\delta g_{\mu\nu}$. That was what we observed when the perturbation was due to conical defect. On the other hand, for the complement of the region $A$, whose size is more than half space, $\alpha>\pi/2$, $\Delta H_{A}\ge\Delta S_{A} $ and the difference is proportional to $T^{\text{bulk}}_{00}$ of the localized source.

\subsection{\label{sec:level1}Thermal perturbation}
In this part, we would like to point out the differences when the perturbation is a mixed state. Although Einstein equations are agnostic whether the source is thermal or pure, in the microscopic description these two cases are substantially different. For example, in the presence of a BH, when the subsystem size reaches a critical value, the difference, $S_{\bar{A}}-S_{A}$ saturates the Araki-Lieb bound namely, $S_{\bar{A}}-S_{A}=S_{A\cup\bar{A}}$ due to homology constraint. In general for a mixed state, one expects separation of the minimal surfaces of a subsystem and its complement which are same in a pure state. Hence the more thermal the system is greater the ignorance becomes. In the thermal case, the amount of information (better to say amount of uncertainty) of the complementary regions of a quantum state does not match due to thermal entropy. Therefore it is not possible to extend the constraints of the previous section to the thermal cases. Let us look at the simple illustration of this fact for a thermal state perturbation on the vacuum,
\begin{align}\label{mixedstateperturbation}
\rho= \frac{|0\ra\la0|+ \sum_{i} e^{-\beta E_{i}} | i \ra \la i |+ ...}{1+ \sum_{i} e^{-\beta E_{i}}+...} .
\end{align}
Here we consider the low tempreture expansion of a thermal state. The CFT side of the story had been studied in \cite{Herzog:2014fra, Cardy:2014jwa}. 

The knowledge that the state is thermal itself is not enough to determine the ontological character of the entropy, as it can be seen as either being entanglement or thermal entropy \cite{Popescu2005}. Surely it is thermal entropy but at the same time one can consider it as the entanglement entropy with its purification\footnotemark[2].\footnotetext[2]{Purification of a mixed state is not unique, yet the mixed state having different purifications have same von Neumann etropy.} Therefore we will not distinguish these two cases, but when we refer the concept as entanglement entropy, it is the entanglement with respect to the purification, not between the complementary regions of the underlying theory.  In other words, we do not refer to the entanglement between subsystem of the mixed state which is very difficult to quantify for an arbitrary mixed state. 

It is a well known fact that entanglement entropy of a system and of its complement do not match for a thermal state. This is also true for the change of entanglement in the leading order. Hence there is no such constraints eq. \eqref{constraints} when the perturbation is a mixed state, which can be shown through non vanishing of full modular hamiltonian. In this case, expectation value of the full modular hamiltonian around the vacuum up to first order does not vanish anymore.
\begin{align}
\delta \la H_{A} \ra= \sum_{i} \tr[(\tr_{\bar{A}}| i\ra\la i | - \tr_{\bar{A}}| 0\ra\la 0 |) H] e^{-\beta E_{i}}
\end{align}
the differences of modular hamiltonians for complementary regions become,
\begin{align}
\delta \la H_{A-\bar{A}}\ra = \sum_{i} \la i | H_{A-\bar{A}} | i \ra e^{-\beta E_{i}}
\end{align}
where $H_{A-\bar{A}} \equiv H_{A}- H_{\bar{A}}$, each of which acts trivially outside its domain. Since $H_{\bar{A}-A}$ does not annihilate excited states, there is no equality between $\delta\la H_{A}\ra$ and $\delta\la H_{\bar{A}} \ra$. Although this relation is simply reflecting the fact that entanglement entropies of complementary regions in a mixed state are different, the first law always satisfied at the linear order, without having any further constraint.

\section{\label{sec:level1}Relative Entropy through Energy and Scale}
At this point, it is clear that the first law-like relations for pure states due to a localized excitation, are actually occurring at the nonlinear level in field variations. In this part we will have a closer look to the origin of the mismatch between $\Delta \la H_{A} \ra$ and $\Delta \la H_{\bar{A}} \ra$ using covariant phase space approach. Although the former equals the change of entanglement entropy in the case of a conical defect perturbation, the latter is always greater than that. So we can begin by asking what the difference $\Delta \la H_{\bar{A}} \ra-\Delta S_{\bar{A}}$ corresponds to in the bulk. Firstly, we will study the difference between $\Delta \la H_{A} \ra$ and $\Delta  S_{A} $ for a state near the vacuum via perturbation theory and later we will generalize the result to any spherically symmetric excitation. 

\subsection{\label{sec:level1}Perturbation theory at the non linear level}
To see the origin of the difference between modular hamiltonian and entanglement entropy, let us study the first law via covariant phase space formulation. We will not give the review of the formulation here rather we will use the fundamental theorem of the formalism which ties the linearized equation of motion in the bulk to change of surface charges. The change of surface charges around the vacuum associated to Rindler horizon generating vector fields match with the information theoretic quantities in the microscopic theory. This identification has been used to derive linearized equation in the bulk through the first law of entanglement entropy in CFT \cite{Faulkner:2013ica}. The fundamental theorem of covariant phase space approach states that,
\begin{align}
d\chi_{\xi} = \omega(\Phi, \delta \Phi, \mathcal{L}_{\xi}\Phi) - 2\xi^{a} \delta E(\Phi)_{ab} \epsilon^{b}
\end{align}
$\Phi$ stands for the whole field content of the theory including gravitational fields. $\omega$ is presymplectic potential and  $\epsilon^{a}$ is $d$ dimensional volume form. $E_{ab}$ is the equation motion derived from full lagrangian including gravitational part. 

The equation above is valid when the equations of motion satisfied for the unperturbed state. $\chi_{\xi}$ is a $(d-1)$ form, whose integral on the boundary region homologous to a minimal surface, yields the change of modular energy and the integral on the minimal surface itself gives the change of the area when the perturbation is considered around the vacuum. 
\begin{align}\label{waldintegral}
\int_{\Sigma} d\chi_{\xi} =  \oint_{\partial\Sigma} \chi_{\xi} = \Delta \la H_{A}\ra - \Delta S_{A} 
\end{align}
where $\Sigma$ denotes the $d$ dimensional timelike hypersurface bounded by minimal surface and infinity.  
Although the theorem is valid for any solution and the perturbations around it. The correspondence between integral of $\chi_{\xi}$ and the information theoretic quantities in the underlying theory is constructed around the vacuum. When $\xi$ is a Killing field, the presymplectic potential vanishes identically, as $\mathcal{L}_{\xi} g=0$.    

The derivation of linearized Einstein equations via the first law of entanglement entropy can be demonstrated simply by the fundamental theorem of the covariant phase space approach, where vanishing of $\Delta H_{A} - \Delta S_{A}$ implies the vanishing of $\int \xi^{a} \delta E_{ab} \epsilon^{b}$.  When the relation is satisfied for every boundary ball, it can be turned into a local equality that is equivalent to linearized Einstein equations. However in our case, we have shown explicitly that the first law of entanglement entropy does not hold. Therefore the difference between modular hamiltonian and entanglement entropy is equal to the linearized equation of motion sourced by the bulk matter stress tensor. If one only considers the gravitational part of the field content, any addition of matter stress can be included at the level of linearized equation as the right hand side of the equality. Hence, the linearized equations are not source free for the perturbed state, which is the bulk point of view on the mismatch between modular energy and entanglement entropy. We should emphasize that we are not deriving linearized Einstein equation  with classical source around the vacuum. The derivation of linearized Einstein equation with source had been proposed in \cite{Swingle:2014uza}. However the source term $\langle T_{\mu\nu}\rangle$ in that derivation was semiclassical by nature hence it can appear as the leading term in the perturbation theory around the vacuum, which corresponds to quantum ($1/N$) corrections on the CFT. Here we assume that the geometry is perturbed by a classical stress energy tensor which appears at the quadratic order in the perturbation theory. Yet one can study the backreaction of the gravitational field on the matter fields through linearized Einstein equations as we will do. It does not seem possible to us that one can derive linearized Einstein equation with a classical source around the vacuum via a perturbation theory on the microscopic state since the source term and gravitational part appear at different orders. To summarize, the difference between the modular Hamiltonian and the entanglement entropy (relative entropy) is equal to modular integral of the bulk stress energy. The exact expression can be obtained simply by  inserting the source term for linearized Einstein equation. 
\begin{align}\label{stressenergycontribution}
S(\rho_A||\rho^{\text{vac.}}_A) = \Delta H_{A} - \Delta S_{A} = \int_{\Sigma} \xi^{a} T_{ab} \,\epsilon^b
\end{align}
which has a simple expression for spherically symmetric configurations, in which case one can evaluate it without detailed knowledge of the energy distribution in the bulk. 
This will be the way we extend the notion of bulk modular energy to excited states beyond perturbation theory. 

The equation \eqref{stressenergycontribution} is obtained by Stokes theorem. In general, variation on the holographic entanglement entropy can originate from two different sources. It can either come as the variation of the minimal surface or variation of the metric field on the surface. To stay in the domain of validity of the Stokes theorem the variation of the entanglement entropy with the minimal surface should vanish. This is the case for linear perturbations since entropy functional (area) is  extremized on the same surface. Hence the perturbation theory should be truncated beyond this order. In the first order, entanglement entropy on the CFT gets contributions only from the expectation value of the boundary stress tensor. At the second order, all the one point functions start to contribute. Let us have a look at how the metric behaves near the boundary in the Fefferman-Graham expansion where the geometry near the boundary can be expanded in the following way,
\begin{align}\label{Fefferman_Graham_expansion}
    g_{\mu\nu} = \frac{L^2}{z^2}(dz^2 + g_{\mu\nu}dx^{\mu}dx^{\nu})
\end{align}
when the asymptotic boundary is flat $g_{\mu\nu}=\eta_{\mu\nu}+\delta g_{\mu\nu}$. In this expansion one can determine the behavior of the scalar fields near the boundary in terms of the one point functions on the CFT, $\phi \sim \gamma z^{\Delta}\langle \mathcal{O}\rangle + ...$ when fed back to Einstein equation with a scalar field, the boundary expansion of the metric is altered in the following way,
\begin{align}
    \delta g_{\mu\nu} \sim a z^{d} \langle T_{\mu\nu}\rangle + b z^{2\Delta}\langle \mathcal{O}\rangle \eta_{\mu\nu}+ ...
\end{align}
where $\Delta$ is the dimension of the operator $\mathcal{O}$. Let us consider a case where both operators contribute at the same energy scale $\mu$ on the boundary theory, then contribution of each term to the entanglement entropy becomes $\langle \mathcal{O}\rangle \sim \mu^{\Delta}$ and $\langle T\rangle \sim \mu^{d}$. The dimensionless perturbation parameter becomes $\mu r \ll 1$ where $r$ is the radius of the sphere CFT lives. Entanglement entropy takes a contribution $(\mu r)^{d}$ from stress tensor at the linear order and $(\mu r)^{2\Delta}$ from the one point functions at the quadratic order. Let us emphasize that the leading order contribution to the entanglement entropy from scalar operators comes at the quadratic level. In this case when the dimension of the operator satisfies $\frac{d}{2}-1<\Delta<\frac{d}{2}$ its contribution becomes the dominant one. In our case, we demand the stress energy contribution to be dominant and truncate the perturbation theory at the quadratic order of the stress energy contribution. In this case, the dimension of the scalar operator can take values between $\frac{d}{2}<\mathcal{O}<d$. Since the perturbation parameter is chosen to be the combination $(\mu r)$ we have control on the entire bulk without restricting ourselves to near boundary regions. This is the weak field limit in the AdS. In the example of the conical defect this corresponds to small angle deficit limit $\delta\theta \ll 1$. In the weak field limit one can decode the matter stress distribution in the entire bulk via the relative entropy on the underlying theory. Mathematically this corresponds to inverse Radon transform. This is the weak field version of the near boundary tomography presented in \cite{Lin:2014hva}.  

\subsection{Appearance of Radial Scale}
In this part we will demonstrate an interesting observation on the contribution of conical defect to the relative entropy. As it is explained in the previous subsection the presence of a localized source increases the relative entropy only in the entanglement wedge that contains it. This is in agreement with the idea of entanglement wedge reconstruction, where one can reconstruct the bulk regions corresponding to the entanglement wedges of those regions. The entanglement wedge reconstruction idea has been studied mostly in cases that exclude backreaction on the geometry  \cite{Dong:2016eik}, although some speculations are made for the case involving backreaction, \cite{Faulkner:2017vdd}. In general, entanglement wedge reconstruction considers the construction of the bulk fields around a classical background using the boundary CFT. Here, we provide further evidence that the conjecture should be valid even when backreaction is considered. The conical defect solution is again a suitable framework where we can explicitly find what the increase is, in the relative entropy due to presence of the defect. Interestingly the contribution of the conical defect can be expressed in terms of the ADM energy of the defect and radius of the region that includes defect or by the radius of the region that can not be probed by the boundary interval that excludes the defect. We have calculated in \eqref{entanglemententropyresultconicaldefect} the change of modular energy for the boundary interval $A$ whose size $\alpha < \pi/2$  in the presence of a small defect. One can easily deduce the corresponding expression for the complementary boundary interval where $\alpha > \pi/2$. 
\begin{align}\label{entanglemententropyresultconicaldefect_complemantary_region}
    \Delta H(\bar{\alpha}) = \Delta H(\pi - \alpha) = 2L\left(1 - (\pi - \alpha)\cot(\pi - \alpha)\right)\Delta M_{\text{ADM}} 
\end{align}
The conical defect perturbation deforms the vacuum into a nearby pure state hence change of the entanglement entropies for the complementary regions should be equal. This allows us to calculate relative entropy for the boundary region $\bar{A}$ directly through the differences of the changes of modular energies of the complementary regions. For a perturbative pure state excitation that is excluded by one of the complementary regions, the first equality below always hold and the second equality is what we have obtained in the case of a conical defect. 
\begin{align}\label{relative_entropy_scale_energy}
    S(\rho_{\bar{A}}||\rho^{\text{vac.}}_{\bar{A}})=\Delta \langle H(\bar{\alpha})\rangle - \Delta \langle H(\alpha)\rangle = 2\pi R  M_{\text{con}}
\end{align}
where $M_{\text{con}}$ is the vacuum subtracted energy of the conical defect. $R$ is the radius of the sphere that, observers having access to region $A$ on the boundary, is blind to. In other words $R$ is the scale in the bulk beyond which, one can not extract any information by having access to the boundary region $A$, (Figure:\ref{bekenstein_bound}). Since the perturbed state is homogeneous or translational invariant, all the regions with same size on the boundary have the same information content which can be denoted by a scale on the boundary. Interestingly there is a one to one map between the scale on the boundary and the radius of the bulk sphere. Boundary observers having access to subsystems of size $2\alpha$ has no information regarding the sphere of radius $R= L \cot\alpha$, where $L$ is the curvature radius of AdS. This expression is remarkable in the sense that it yields the information theoretic content of the bulk excitations in a non-local way in terms of the bulk quantities. A similar expression is used in, \cite{Verlinde:2016toy} to motivate the information theoretic effect of introducing matter onto spacetime. This effect appears as a reduction of entanglement entropy once it is postulated that surface area is the measure for entanglement entropy of the quantum state describing spacetime. In the next section we will give the derivation of the radial scale $R$ in higher dimensional generalization of conical defects where excited state can be considered as a perturbation over the vacuum. 
\subsection{\label{sec:level1}Higher Dimensional Generalizations for Perturbative \\Excitations}
\begin{figure}[t!]\label{figure:coordinatesystem}
	\centering
	\def\svgwidth{0.5\textwidth}
\begingroup%
  \makeatletter%
  \providecommand\color[2][]{%
    \errmessage{(Inkscape) Color is used for the text in Inkscape, but the package 'color.sty' is not loaded}%
    \renewcommand\color[2][]{}%
  }%
  \providecommand\transparent[1]{%
    \errmessage{(Inkscape) Transparency is used (non-zero) for the text in Inkscape, but the package 'transparent.sty' is not loaded}%
    \renewcommand\transparent[1]{}%
  }%
  \providecommand\rotatebox[2]{#2}%
  \newcommand*\fsize{\dimexpr\f@size pt\relax}%
  \newcommand*\lineheight[1]{\fontsize{\fsize}{#1\fsize}\selectfont}%
  \ifx\svgwidth\undefined%
    \setlength{\unitlength}{196.59705624bp}%
    \ifx\svgscale\undefined%
      \relax%
    \else%
      \setlength{\unitlength}{\unitlength * \real{\svgscale}}%
    \fi%
  \else%
    \setlength{\unitlength}{\svgwidth}%
  \fi%
  \global\let\svgwidth\undefined%
  \global\let\svgscale\undefined%
  \makeatother%
  \begin{picture}(1,1.06274097)%
    \lineheight{1}%
    \setlength\tabcolsep{0pt}%
    \put(0,0){\includegraphics[width=\unitlength]{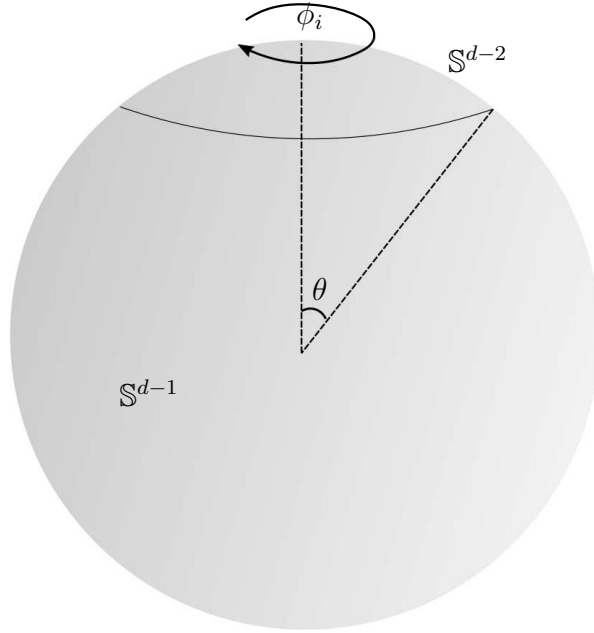}}%
    \put(0.51149751,0.55780458){\color[rgb]{0,0,0}\makebox(0,0)[lt]{\lineheight{1.25}\smash{\begin{tabular}[t]{l}$\theta$\end{tabular}}}}%
    \put(0.7411191,0.94316871){\color[rgb]{0,0,0}\makebox(0,0)[lt]{\lineheight{1.25}\smash{\begin{tabular}[t]{l}$\mathbb{S}^{d-2}$\end{tabular}}}}%
    \put(0.18620338,0.37861043){\color[rgb]{0,0,0}\makebox(0,0)[lt]{\lineheight{1.25}\smash{\begin{tabular}[t]{l}$\mathbb{S}^{d-1}$\end{tabular}}}}%
    \put(0.48461843,1.02409514){\color[rgb]{0,0,0}\makebox(0,0)[lt]{\lineheight{1.25}\smash{\begin{tabular}[t]{l}$\phi_i$\end{tabular}}}}%
  \end{picture}%
\endgroup%

	\caption{Coordinate system used for constant time slice of AdS$_{d+1}$. Opening angle $2\alpha$ of the boundary ball $\mathbb{S}^{d-2}$ is measured by the coordinate variable $\theta$.}
\end{figure}

In this section we will extend the connection between relative entropy and the bulk modular energy to higher dimensions for spherically symmetric perturbative excitations. We will show how the bulk radial scale enter into the calculation which will later be used for the derivation of the Bekenstein bound in the bulk. $d+1$ AdS can be represented by the hyperboloid,
\begin{align}\label{hyperboloid_equation}
    X_{0}^{2}+X_{d+1}^{2}-\sum_{i=1}^{d} X_{i}^{2}=L^2
\end{align}
embedded in $d+2$ dimensional flat space. Equation \eqref{hyperboloid_equation} can be solved by setting,
\begin{align}
    X_0 &= L \cosh\chi \cos\tau,\nonumber\\
    X_{d+1} &= L \cosh\chi \sin\tau,\nonumber\\
    X_{i} &= L \sinh\chi\, \Omega_i,\nonumber
\end{align}
where $\sum_{i=1}^{d-1} \Omega_i =1$ and spans the trigonometric functions of $\theta, \phi_i$ where $i$ runs in $\{1...d-2\}$. The solution $(\chi\ge 0,\,\, 0\le\tau\le2\pi)$ covers the entire hyperboloid hence yields the global description of AdS$_{d+1}$, whose metric becomes,
\begin{align}\label{hyperbolicslicing}
ds^2=L^2(-\cosh^2\chi d\tau^2 + d\chi^2+\sinh^2\chi d\Omega^2_{d-1})\,
\end{align}
We would like to compactify the solution such that boundary resides at a finite value of the radial direction. The casual structure of AdS$_{d+1}$ can be studied by the following coordinate transformation,
\begin{align}
    \sinh\chi = \tan\rho,\hspace{1cm}0\le \rho\le\pi/2,
\end{align}
the metric becomes,
\begin{align}\label{compactified_global_coordinates}
    ds^2 = \frac{L^2}{\cos\rho^2}\Big(-d\tau^2+d\rho^2+\sin^2\rho \underbrace{(d\theta^2+\sin^{d-2}\theta\, d\Omega^2_{d-2})}_{d\Omega^2_{d-1}}\Big)
\end{align}
In this coordinate \ref{figure:coordinatesystem} system boundary is located at $\rho=\pi/2$. Size of the $\mathbb{S}^{d-2}$ boundary ball, $A$ is determined by the coordinate variable $\theta$. The Killing field that generates the Rindler horizon in the global coordinates is given by,
\begin{align}\label{horizon_generating_killing_field}
    \xi^{a} = \frac{\cos\tau\sin\rho\cos\theta - \cos\alpha}{\sin\alpha}\partial_{\tau}+\frac{\sin\tau\cos\rho\cos\theta}{\sin\alpha}\partial_{\rho}-\frac{\sin\tau\sin\theta}{\sin\rho\sin\alpha}\partial_{\theta}
\end{align}

$\xi^{a}$ vanishes on the minimal surface that is homologous to the boundary ball $A$ and it has unit surface gravity. $\alpha$ is the size of the boundary ball. Let us turn back to the problem of generalizing out observation on the conical defect to higher dimensions. Consider a spherically symmetric perturbative excitation in the AdS$_{d+1}$. Stress energy distribution characterizing this excitation depends only on the radial coordinate $T^{\text{bulk}}_{ab}\equiv T^{\text{bulk}}_{ab}(\rho)$ which  corresponds to a uniform (spherically symmetric) energy density on the boundary. Let us pick the constant time slice to study the relative entropy around the vacuum whose geometric dual is given by the solution \eqref{compactified_global_coordinates}. Suppose there is no excitation in the casual wedge of the boundary region $A$ $\ie$ all of the bulk excitation confined inside the complement $\bar{A}$. In this case the difference of the change of modular energies equal to the relative entropy of the complement and can be expressed as the modular integral of the bulk stress energy.
\begin{align}\label{bulk_integral_for_relative_entropy}
    S(\rho_{\bar{A}}||\rho^{\text{vac.}}_{\bar{A}})=\Delta \langle H(\bar{\alpha}) - H(\alpha)\rangle &=2\pi L \int_{\rho\le\rho_0} \frac{\sin\rho\cos\theta - \cos\alpha}{\sin\alpha} T^{\text{bulk}}(\rho)\,d^{d}V\nonumber\\
    &= 2\pi L \cot\alpha\, \Delta M_{\text{ADM}}
\end{align}
 Note that the expression is given in terms of curvature radius of the AdS$_{d+1}$ as a necessity of dimensionless nature of relative entropy. The appearance of curvature scale plays an important role in the identification with the radial coordinates in the bulk. In the expression above $\rho_0$ represents the deepest point that can be probed via the boundary region $A$.  Observers that have access to the smaller boundary can not decode the bulk beyond this point, hence it denotes a sphere of ignorance. Let us look at the physical interpretation of the factor $\cot\alpha$ from the bulk point of view. The equation of the minimal surface that are homologous to $(d-2)$ spheres on the boundary is given by,
\begin{align}\label{equation_for_surface}
    \sin\rho\cos(\theta-\theta_0)=\cos\alpha
\end{align}
$\theta_0$ denotes the center of the boundary ball in $\theta$. The deepest point that the surface reach has the angular coordinate $\theta = \theta_0$. The radial coordinate of the point of ignorance becomes, $\sin{\rho}=\cos\alpha$. Let us represent the the radius of the the sphere of ignorance using the spherical coordinates. The radial coordinate sits in front of the angular directions in spherical coordinates as $R^2 d\Omega^2$. In the global AdS this corresponds to $\tan\rho$. Using the expression for the location of the tip in terms of $\alpha$, we infer that radius of the sphere that observers having access to region $A$ can not access becomes,
\begin{align}\label{radial_scale_in_the_bulk}
    R_{scale}=L\cot\alpha
\end{align}
Therefore we have derived that in the perturbative regime the modular energy of spherically symmetric excitation can be seen as a non local contribution that depends on the size of the system. A similar result is used in \cite{Verlinde:2016toy} to motivate the idea that matter reduces the entanglement entropy of the spacetime in a way that is proportional to the radial scale of the hypothetical box that contains the excitation. Our result also indicate that it should be possible to construct modular modular hamiltonian for a spherical region in the bulk. To sum up we have obtained the following expression for the modular energy contribution of the bulk excitation to the entanglement wedge that includes the excitation,
\begin{align}\label{modular_energy_in_the_bulk}
    \Delta\mathcal{E}^{\text{bulk}} \equiv 2\pi R_{scale} \Delta M_{\text{ADM}}
\end{align}
where bulk modular energy contribution to the entanglement wedge is defined as  $\Delta\mathcal{E}^{\text{bulk}}\equiv \Delta\langle H_{\bar{A}}- H_A \rangle$. In the next section we will derive this relation completely through underlying theory and we will put some emphasis on the differences between pure and thermal state excitations.

\section{\label{sec:level1}Bekenstein bound and AdS/CFT}
The Bekenstein bound, \cite{Bekenstein:1980jp, Bekenstein:2003dt} is a limit on the entropy that can be contained in a physical system or object with a given size and total energy. The heuristic yet deep derivation of the bound employes the black hole thermodynamics together with \textit{generalized second law}(GSL). Generalized second law is the extension of the the second law of thermodynamics to the systems involving black holes. The law states that the environment and black hole system together evolves in such a way that total entropy of the combined system does not decrease. Bekenstein derived the bound mainly employing this principle.

Let us have a look at the following \textit{gedanken} experiment. A composite system of radius $R$ with total energy $E$ and entropy $S_{\text{box}}$, falls into a Schwarzschild black hole black hole of mass, $M$ where $M\gg E/c^2$, such that temperature of the black hole stays same in the process. Suppose the system is dropped from a large distance, such that before it becomes part of the black hole, equal amount of entropy is radiated by the black hole. Hence at the end of the process black hole mass stays the same and therefore black hole entropy does not change. Given the process is reversible, the radiation entropy becomes $E/T_{BH}.$\footnotemark[3]\footnotetext[3]{Taking curvature into account may alter the amount of entropy emitted by thermal radiation, however these are $\mathcal{O}$(1) effects that can be absorbed into the coefficients in the bound.} Thus the overall change in the entropy of the universe becomes,
\begin{align}
\Delta S = E/T_{BH} - S_{\text{box}}
\end{align}  
One can choose the Schwarzschild radius larger than then the size of the box $R$, such that system will fall into black hole without being torn apart, $\ie$ the relation between size of the box and Schwarzschild radius is controlled by an $\mathcal{O}(1)$ parameter $\lambda$, $R_{\text{BH}} \sim \lambda R$. Since GSL implies $\Delta S_{\text{universe}} \ge 0$, one puts a bound on the total entropy that can be contained in the box.
\begin{align}
S_{\text{box}} \le \lambda R E /c\hbar
\end{align}
What happens to the bound when the number of species in the system increase was a long standing puzzle. One alternative and more elaborate derivation of the bound has been given in  \cite{Casini:2008cr} which yields some understanding on how bound preserves its validity when the number of species is increased. The quantum information theoretic derivation of the bound is based on the positivity of relative entropy and has been derived for ball shaped regions in the CFT for the excitations around the vacuum density matrix and for QFT s around the density matrix corresponding the to thermal state in the Rindler space. The derivation is given for few cases where local expression for the modular Hamiltonian is known. Since the manifestation of the bulk version of the Bekenstein bound in the underlying theory employing similar elements, let us briefly reproduce derivation of the bound in QFT through positivity of relative entropy. The derivation of the bound using the modular Hamiltonian of the ball shaped region in the CFT is more illuminating as it yields a natural definition for the system of size $R$. Let us consider this system as the ball shaped region itself. The positivity of relative entropy implies, $S(\rho|\rho_{\text{vac.}})\ge 0\implies \Delta S_{\text{box}}\le \Delta\langle H\rangle$ where $\Delta$ represents the vacuum subtracted quantities. Inserting the local expression for modular Hamiltonian in general dimensions,
\begin{align}\label{bekenstein_bound_positivity}
    \Delta S_{\text{box}}\le 2\pi\int^{R}_{0}dr\, r^{d-1}\int d\Omega_{d-1} \frac{R^2 - r^2}{2R} \Delta \langle \hat{T}_{00}\rangle
\end{align}
For spherically symmetric distributions one can take the integral and turn the local integral over the energy density into a total energy relation. For example for a localized source at the center one obtains $\Delta S_{\text{box}}\le \pi R\, \Delta E$ and for a uniform energy distribution $\Delta S_{\text{box}}\le \pi R\, \Delta E/(d+2)$, both of which satisfies the bound up to an order one factor. 
In the original derivation of the bound, gravity plays a central role, yet the expression is independent of $\GN$. On the other hand, the derivation of the bound based on the positivity of relative entropy is completely quantum mechanical hence explains why the bound is independent of $\GN$. Although Bekenstein bound is independent of $\GN$ it exists for gravitational system even when the self gravitation is strong. Saturation of the bound for black holes is a nice example of this situation. In this case size of the box becomes Schwarzschild radius, which is the radial coordinate rather than the geodesic distance. Since the information theoretic derivation of the bound exploits vacuum subtracted quantities, size of the box becomes ambiguous when system has back-reaction on the geometry. One needs to find a reference manifold with respect to which, the size of the box is defined. As we will show size of the system is fixed with respect to the AdS$_{d+1}$ in the formulation of Bekenstein bound in the bulk. 


\subsection{\label{sec:level1}Thermal state excitations}
In the derivation of the expression \eqref{modular_energy_in_the_bulk}, we used explicitly that the excited state is a pure state, in which the change of entanglement entropies of the complementary regions are equal. The purity of the excited state is used to relate the $\Delta \la H_{\bar{A}} - H_{A} \ra$ to the bulk modular integral of the stress energy tensor. To be explicit we employed the equality of $\Delta S_{\bar{A}}$ and $\Delta \la H_{A} \ra$ through their relation to $\Delta S_{A}$. In the case of a thermal state, the change of entanglement entropies would not be equal anymore. Therefore, we could not derive the same expression when the excitation is a mixed state (\ref{mixedstateperturbation}). In that case,  bulk excitation would carry information that can not be deduced from the underlying state without access to auxiliary purification.

On the other hand, in the macroscopic description of the so-called first law, we have only specified change in the expectation value of the boundary stress tensor  $\Delta \la  T_{\mu\nu}\ra $ together with the knowledge of the purity of the perturbed state. The change in the expectation value of the boundary stress energy alone, does not specify the microscopic nature of the perturbation. Pure and mixed state perturbations are quite different although they may yield equal amount of change in the energy of the system. One important difference in their nature, as we have explained in section 3: a pure state perturbation with a non vanishing net energy increase can not take place at the linear level while that for a thermal state can. This is a very restrictive statement which implies that linearized Einstein equations with a classical source can not take place at the linear level from the point of microscopic theory, which as we have explained, took place at non linear level and source should be considered as the back-reaction of geometry on the stress energy tensor. This is expected, since bulk stress energy tensor vanishes for the perturbations at the linear level  around the vacuum. However when the perturbation is a mixed state then there is no such constraint on the change of total  energy of the system. To sum up when one only specifies the change of boundary stress tensor, one does not know the information theoretic content of the perturbation. A mixed state perturbation and a pure state one only differs in terms of their entanglement entropic content. 

\subsubsection{\label{sec:level1}Thermal states at the linear level}
Before observing the manifestation of the Bekenstein bound in the microscopic theory as the monotonicity of relative entropy on the CFT, let us study thermal states that are perturbations around the vacuum $\rho=\rho_{0}+\delta\rho$ to see how different the observable $\delta S_{\bar{A}-A}\equiv\delta S_{\bar{A}}-\delta S_{A}$ behaves, which was vanishing for any pure state excitation. 

In the case of a pure state, the change of entanglement for some region and its complement is equal, $\Delta S_A=\Delta S_{\bar{A}}$. Conical defect perturbation was an example of this case, where the defect was included only in one of the entanglement wedges, yet the information content of the complement is the same. We interpret this as by arguing that defect does not carry entropy in itself with respect to the underlying state containing it. If the perturbation is in the form of a thermal state, then one would expect totally different behavior. Note that we were also not allowed by the constraints, \eqref{constraints}, to study the first law of entanglement for pure states which has $\delta E\neq 0$, which is not the case for thermal states.

This is simply because the perturbation carries entropy in the form of entanglement with its purification. Let us consider a mixed state perturbation at the linear order. It satisfies the first law of entanglement entropy, both for the boundary region $A$ and its complement $\bar{A}$.  Using the first law of entanglement, we can find the difference $\delta S_{\bar{A}}-\delta S_{A}$ which is considered as a measure of information associated to the entanglement wedge of $\bar{A}$, that can not be retrieved from $A$.  
 \begin{align}
\delta S_{\bar{A}}-\delta S_{A}=\delta \la H_{\bar{A}}\ra-\la\delta H_{A}\ra
\end{align}
The change of entanglement entropy of the underlying state can be decomposed into two contributions. The area contribution and entanglement entropy of quantum fields in the bulk which emerges as quantum corrections to the underlying state. As we have explained in great detail in section 3, linearized perturbations on the underlying state can only change the total energy of the state if they are of thermal nature.
\begin{align}
   \delta S_{\bar{A}} = \frac{\delta A}{4\GN} + \delta S^{\text{bulk}}_{\Sigma_{\bar{A}}} 
\end{align}
where $\Sigma_{A}$ denotes the entanglement wedge of $A$. The contribution from bulk fields can be expressed as a local integral expression in the linear level \cite{Leichenauer:2015nxa}.
\begin{align}
\delta S^{\text{bulk}}_{\Sigma_{A}} = \int_{\Sigma_{A}} \zeta^{\mu} \langle T_{\mu\nu}\rangle \,d\Sigma^{\nu}
\end{align}
when all the contribution is confined into the sphere of ignorance one can equate the bulk entanglement contribution of this region to the difference of change of modular energies. Because in this case even the perturbation is of thermal nature, the change of areas would be equal due to extremal character of the surfaces. Hence,
\begin{align}\label{linearthermal}
\delta S^{\text{bulk}}_{\mathbb{S}_{R}} =  2\pi R \delta M_{\text{ADM}}
\end{align}
the bulk entanglement entropy resides in the sphere of ignorance is defined as
\begin{align}
    \delta S^{\text{bulk}}_{\mathbb{S}_{R}} = \int \Theta(R - r) \zeta^{\mu} \delta \langle T_{\mu\nu}\rangle \,d\Sigma^{\nu}
\end{align}
we obtain the entropic version of the \eqref{radial_scale_in_the_bulk}.
where $R=L \cot \alpha$ and $\alpha$ is the angular radius of the boundary region $A$. This is the maximum entropy that can be contained in the spherical region around the origin with radius, $R$ for a system with energy $M_{\text{ADM}}$. It shows us that, the difference between entanglement entropies of complementary regions in a  thermal perturbative excitation at the linear level is equal to saturation of Bekenstein bound for a system with energy $M_{\text{ADM}}$ and size $R$. Indeed this is the region that the observer who has control on system $A$ is blind to. Any deviation from thermal nature (mixture of thermal state and pure state as an ensemble), decreases the $\delta S_{\bar{A}-A}$ as it is zero in pure state.

Note that the expression diverges in the limit $\alpha \rightarrow 0$. In this case region $\bar{A}$ covers the whole boundary. How could we make sense of this expression, in this limit? There are two scales in the problem: that of $\alpha$ and $\beta$, the inverse temperature of the system. Although \eqref{linearthermal} does not contain $\beta$ explicitly, it is absorbed into $\delta E_{\text{CFT}}$ \eqref{mass_energy_identification}, which should be read in terms of the energy of lowest excited state weighted with the Boltzman constant and the degeneracy of the state  $\delta E_{\text{CFT}} = \sum_i g_i  E_{i}  e^{-\beta E_{i}}$. This expression is sensitive to the order of limits and to make sense of it, one should consider the limit $\beta\rightarrow\infty$ before $\alpha\rightarrow \pi$ \cite{Cardy:2014jwa}. Although, order of limits can let us make sense of the expression in $\alpha \rightarrow 0$ limit, it is still an open question, at least to the author, how can we make sense of expression as an operator expression, since same expression can also be used to evaluate $\Delta \la H_{\bar{A}}\ra$ for non perturbative excited states.

\subsection{\label{sec:level1} Bekenstein Bound in the Bulk}
Until now we have carried out a perturbative analysis using covariant phase space formulation. We computed the $\Delta \la H_{\bar{A}-A}\ra $ and $\delta S_{\bar{A}-A}$ as an integral of the bulk stress energy tensor using the fundamental theorem of the covariant phase space formalism. To emphasize again this is only valid when the geometric dual of the excited state can be seen as perturbation of the metric field around the AdS. In this section, we will show that one can go beyond perturbation theory solely using CFT modular Hamiltonian. Although the difference $\Delta \la H_{\bar{A}} - H_{A} \ra $ can be obtained only referring to the boundary quantities for non perturbative excitation, it has the similar bulk interpretations. Again the quantities involved in the expressions have different nature for pure and mixed states.

The modular Hamiltonian for a ball shaped region for the vacuum of the CFT is an operator expression \eqref{d-dimensionalmodularhamiltonian}. Hence it has no restriction on the state that the operator is evaluated. Until now we have used this expression for the states that are close to the vacuum, in that case there exists a dual bulk expression for the relative entropy when one of the entanglement wedges excludes any excitation in the bulk. 

Let us consider an excited state $|\Psi\rangle$ which is orthogonal to the vacuum $\langle \Psi|0\rangle = 0 $. One can study change in the modular energies by taking the expectation value of the stress energy around $|\Psi\rangle$. Considering the differences of the changes for complementary regions we obtain,
\begin{align}\label{full_modular_energy}
    \Delta \la H_{\bar{A}-A} \ra = 2\pi\int^{\pi}_{0}r^{d-1} \,d\Omega_{d-2} \sin^{d-2}\theta \left( r \frac{\cos\theta-\cos\alpha}{\sin\alpha}\right) \Delta\la \hat{T}_{00}\ra(\vec{r})
\end{align}
where $\Delta\la\hat{T}_{00}\ra\equiv \la\hat{T}_{00}\ra_{\Psi} - \la\hat{T}_{00}\ra_0$. For a pure state $\Delta\la H_{\bar{A}-A}\ra = S(\rho_{\bar{A}}|\rho^{0}_{\bar{A}}) - S(\rho_{A}|\rho^{0}_{A})$. In this case one can not associate any entropy to the bulk excitation in the form of entanglement. Let us evaluate \eqref{full_modular_energy} for homogenous excitation where energy density of the state is given as $\epsilon = \frac{\Delta E}{r^{d-1}\Omega_{d-1}}$, in this case bulk dual of the state becomes spherically symmetric.

\begin{align}
    \Delta \la H_{\bar{A}-A} \ra &= 2\pi r \Delta E\, \frac{\text{Vol}(S^{d-2})}{\text{Vol}(S^{d-1})}\int^{\pi}_{0} \, \sin^{d-2}\theta \left( \frac{\cos\theta-\cos\alpha}{\sin\alpha}\right)\nonumber\\
    &=2\pi r \cot\alpha\, \Delta E_{\text{CFT}} .
\end{align}
The expression is the higher dimensional non-pertubative generalization of the one that is obtained in \eqref{relative_entropy_scale_energy} for conical defects. The expression is remarkable as it is valid in any dimension yet it is more interesting when one understand it in terms of the bulk quantities. Let us elaborate the identification with the bulk in more detail. We choose the cylindrical description $\mathbb{R} \times \mathbb{S}^{d-1}$ of AdS$_{d+1}$. The metric of the AdS$_{d+1}$ is given in \eqref{compactified_global_coordinates}. Although one can fit the geometry on a finite piece of paper, the actual radius of the boundary sphere becomes infinite. However this is an overall conformal factor that can be removed such that volume of the boundary sphere becomes finite. Hence flow of time in the bulk and boundary descriptions are different as it is measured by the lapse function $N$ (=$\sqrt{g_{tt}}$) in the ADM description. In the global coordinates $N\rightarrow R/L $ asymptotically.\footnotemark[4] \footnotetext[4]{The identification outlined above appears in  a more rigorous way in the duality between AdS$_5$ $\times \mathbb{S}^{5}$ and $SU(N)$  Yang-Mills. The relation  $L^3/\GN=2N^{2}/\pi$  makes it possible to relate the mass to the dimension of the gauge group $N$. Then $r E_{\text{casimir}}   = L M_{\text{casimir}}  = 3(N^2 -1)/16 $.}  The energy of the state in the CFT is proportional to the mass of the dual gravitational solution,
\begin{align}\label{mass_energy_identification}
   \Delta M_{\text{ADM}} = \frac{r}{L} \Delta E_{\text{CFT}}
\end{align}
where $r$ is the curvature radius of $\mathbb{S}^{d-1}$ while $L$ is the curvature of the AdS. Equality above also ensures a dimensionless identification in the information theoretic observables. If one identifies these two scales then energies of the theories are naturally identified. This identification is necessary to recognize the radial scale of the deepest point that can be probed by the state of $A$ ($\text{Vol}(\bar{A})\ge \text{Vol}(A)$) in CFT. Remember that the deepest point in the bulk that can be reached via the boundary region $A$ was given in \eqref{radial_scale_in_the_bulk}. Inserting 
these we see that the change in the full modular hamiltonian in the complementary regions becomes,
\begin{align}
    \Delta \la H_{\bar{A}-A} \ra = 2\pi R_{scale} \Delta M_{\text{ADM}}
\end{align}
This is entirely a bulk expression due to the natural identification between modular energies of the CFT and the gravitational dual. Once again, just like the perturbative case \eqref{bulk_integral_for_relative_entropy}, we have observed that for spherically symmetric excitation, the difference in the entanglement energies of the complementary states of the underlying theory have an expression in terms of bulk quantities.  

It has been emphasized along the paper that when excitation is pure state, the change of entanglement for complementary regions are equal $\Delta S_{\bar{A}-A}=0$. In this case one can not associate extra new correlation in the form of entanglement to the bulk excitation for scales less then the radial probing point of the boundary observes having access to $A$. In the language of bit threads \cite{Freedman:2016zud} no additional thread (additional in the sense that comes by the excitation on top of the vacuum tread configuration) ends up in the bulk. Purity of the state constrains the amount of information that is missing beyond the scale $R$. On the other hand for a thermal state $\Delta S_{\bar{A}-A}\ne 0$. Let us focus to spherically symmetric thermal excitations again. For non-perturbative excitations we do not have the localized 
\begin{figure}[t!]
    \centering
    \includegraphics[width=\textwidth]{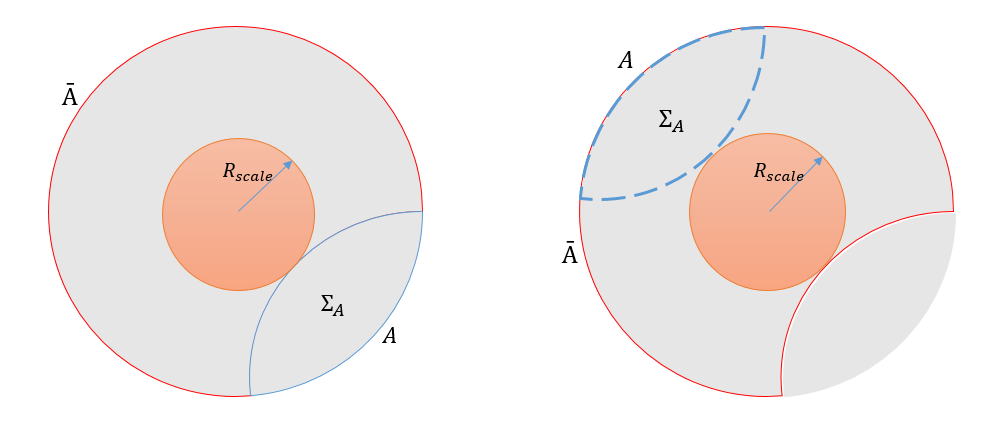}
    \caption{The difference between change of entanglement entropies of the complementary regions in CFT$_d$ for a thermal state with respect to vacuum is bounded by the maximum entropy that can be contained in the region excluded by the smaller interval. Spherical symmetry allows one to translate the region $A$, which is pictured on the right. Monotonicity of relative entropy puts a bound on the difference of entanglement entropies, $\Delta S_{\bar{A}-A}\leq 2\pi R_{scale} \Delta M_{\text{ADM}}$, which is the consequence of Bekenstein bound in the geometric description.}
    \label{monotonicity_of_relative_entropy}
\end{figure}
expression of the bulk stress tensor anywhere in the bulk which was possible for states that are dual to geometries that can be expressed as perturbations of the metric around the AdS. Positivity of relative entropy dictates that,
\begin{align}
S(\rho_{A}||\rho_{0 A})\ge 0,\hspace{1cm}\implies \hspace{1cm}\Delta \la H_{A}\ra-\Delta S_{A}\ge 0
\end{align}
Since, excitation is not perturbation around the vacuum, one can not equate the right hand side of the first equation to the modular integral of bulk stress energy. Possibly it is not even in the form of local expression. However assuming spherically symmetry for excited state one can understand the implications for the bulk physics in the non perturbative level. As a consequence of spherical symmetry one can translate the region $A$ such that $\bar{A}\supseteq A$ without altering $\Delta S_{A}$ or $\Delta \la \hat{H}_{A}\ra$. Following rotations, Further impose the monotonicity of the relative entropy,
\begin{align}
S(\rho_{\bar{A}}||\rho_{0 \bar{A}})\ge S(\rho_{A}||\rho_{0 A}) \hspace{1cm}\implies \hspace{1cm} \Delta S_{\bar{A}-A} \leq \Delta \la H_{\bar{A}}\ra-\Delta \la H_{A}\ra .
\end{align}
We have already calculated right hand side using the boundary expressions of the modular hamiltonian. 
\begin{align}\label{bekenstein_bound_final}
|\Delta S_{\bar{A}-A}| \le 2\pi R_{\text{scale}} \Delta M_{\text{ADM}} \, .
\end{align}
The inequality is universal in the sense that it is independent of the details of the excitation, and how it is organized spatially in the gravitational theory apart from its spherical symmetry. Remember that we have encountered necessity of spherical symmetry in the derivation of the Bekenstein bound using positivity of relative entropy \eqref{bekenstein_bound_positivity} also in the QFT \cite{Casini:2008cr}. One can deviate from spherical symmetry by considering $\mathcal{O}(1)$ deformations of the bound. Using the symmetry we argue that the entropy contained in the sphere of ignorance is bounded by the difference of vacuum subtracted entropies of the complementary regions on the boundary. 
\begin{align}\label{bulkboundaryentropyconjecture}
	\Delta S^{\text{bulk}}_{\mathbb{S}_{R}}\le \Delta S_{\bar{A}-A}
\end{align}
Remember that $\Delta S_{\bar{A}-A}\equiv \Delta S_{\bar{A}}-\Delta S_{A}$. The bound becomes and equality in the perturbative limit as shown in eq. \eqref{linearthermal}. 
The symmetry of the system actually reduces the effective dimensions to one and allows us to represent entanglement entropy using the two scales of the system, namely energy and size of the box. In this one effective spatial dimensional information space, the difference between entanglement entopies of the complementary regions in the microscopic system, is bounded by the Bekenstein bound for a system with radius $R$ and energy $M$ in the bulk, which is indeed the region that is excluded by any observation on $\text{scale}(A)$. If all the energy was contained in the radius $r_{\text{bulk}}< R_{\text{scale}}$ and organized to in a way to saturate the Bekenstein bound, then $|\Delta S_{\bar{A}-A}| = 2\pi R_{\text{scale}} \Delta M_{\text{ADM}}$. The inequality also holds perturbatively, at the linear order as we have shown. 
In the transition from a pure state to a thermal state, $|\delta S_{\bar{A}}-\delta S_{A}|$ is interpolates between $0$ and $2\pi R_{\text{scale}} \Delta M_{\text{ADM}}$. Whenever some of the thermal energy is replaced by an equal amount of energy corresponding to a pure state, the difference between entanglement entropies decreases.

On the other hand, we should be careful using $\hat{H}_{A}$ on thermal states \footnotetext[5]{Phase transitions on the modular hamiltonian takes place also for the disjoint intervals depending on the distance of separation. Another example is the entanglement entropy in the conical defect geometry when it is seen as the entanglement entropy of the disjoint intervals in the parent theory. In both of these cases, the phase transition on entanglement entropy is due to a jump in the saddle point and mutual information between disjoint intervals is a probe of different phases.}  when the bulk dual is a black hole solution. In this case one observes a phase transitions \footnotemark[5] in the entanglement entropy along the continuous increment of the system size. These phase transitions are formulated as homology constraints for the minimal surfaces in the bulk \cite{Hubeny:2013gta}.  Therefore the local expression of $\hat{H}_{A}$ is not valid for regions bigger than the critical size $\theta_{\text{critical}}$ beyond which phase transitions take place as formulated in homology constraints. The point where phase transition took place manifest itself as a sudden jump on the minimal surface. This is also the point where Araki-Lieb bound is saturated. In the next section we will study the relation between Araki-Lieb bound and the one we have derived via monotonicity of relative entropy.

\subsection{Comparison with Araki-Lieb bound}
The bound we have derived in \eqref{bekenstein_bound_final} using the monotonicity and positivity of relative entropy for certain class of excitations has the same quantity with Araki-Lieb bound on the left hand side of the inequality. It is an interesting exercise to study these two inequalities together and see whether the bound derived here is trivial when it is compared to Araki-Lieb bound. 

The notion of entanglement entropy we had been referring along the paper is von Neumann entropy, which quantifies the extent to which the state represented by $\rho$ fails to be a pure state. The reason that von Neumann entropy serves as entanglement entropy is that when the state $\rho$ is obtained from a pure state by tracing over part of the Hilbert space representing a subsystem, such as the one that can not be accessed by the observer, then von Neumann entropy measures the entanglement entropy between subsystem that is traced out and the rest. Suppose the Hilbert space of the full system $\mathcal{H}_{full}$ factorizes into Hilbert space of two subsystems, $\mathcal{H}_{full}=\mathcal{H}_{A}\otimes \mathcal{H}_{\bar{A}}$. Then for each subsystem we have corresponding density matrices defined by tracing over the complementary subsystem $\rho_{A,\bar{A}}=\text{Tr}_{\bar{A}, A}(\rho_{\text{full}})$. The entanglement entropies that are associated to each density matrix can be shown to satisfy following inequalities \cite{Araki1970aheh},
\begin{align}
|S(\rho_{A})-S(\rho_{\bar{A}})|\le S(\rho_{\text{full}}) \le S(\rho_{A}) + S(\rho_{\bar{A}}) .
\end{align}
The first part of the triangle inequality is usually referred as Araki-Lieb bound, while the second is known as subadditivity. The Araki-Lieb bound is derived from subadditivity. We have also derived an inequality that is similar to the first part of the triangle inequality. We have observed that the difference of entanglement entropies of the complementary subsystems follows the Bekenstein bound given in terms of the bulk quantities. Our bound becomes non trivial compared to Araki-Lieb bound when $2\pi M R \le S(\rho_{\text{full}})$. This happens when system sizes on the CFT approach each other. In this limit, Bekenstein bound takes over the Araki-Lieb. Let us compare these two bounds by considering $d$ dimensional CFT at finite temperature and having geometric dual as AdS-Schwarzschild black hole.

The metric for $(d+1)$-dimensional static solution for asymptotically AdS spacetimes is given by,

\begin{align}
    ds^2 = -f(r) dt^2 + \frac{dr^2}{f(r)} + r^2 d\Omega^{2}_{d-1}, \hspace{1cm} f(r) = 1+\frac{r^2}{L^2}-\frac{\mu}{r^{d-2}}
\end{align}
where $\mu = \frac{16\pi \GN M}{\Omega_{d-1} (d-1)}$.
On this state Araki-Lieb bound can be expressed in terms of the black hole entropy.
\begin{align}
    |\Delta S_{A-\bar{A}}|(\alpha)\le \frac{A(r_{+})}{4\GN} = \frac{r_{+}^{d-1}\Omega_{d-1}}{4\GN}
\end{align}
where  $ \Delta S_{A-\bar{A}} = |\Delta S(\rho_{A})-\Delta S(\rho_{\bar{A}})|$ and $r_{+}$ is the largest solution to the equation,
\begin{align}
    1+\frac{r^2}{L^2}-\frac{\mu}{r^{d-2}}=0.
\end{align}
We considered vacuum subtracted quantities on the right hand side. Since the vacuum entanglement entropies of the complementary states are equal, this subtraction does not change the difference. Since the solution is spherically symmetric we can use it to set Bekenstein bound. Let us calculate the mass of the solution in terms of the $r_{+}$. The calculation had been carried first by Hawking-Page \cite{Hawking:1982dh} in $d=3$ and later generalized to arbitrary dimensions by Witten \cite{Witten:1998zw}. The key point is the connection between action and the partition function $I = -\log Z$. Energy of the excitation can be calculated by change of the action with respect to inverse temperature $E = \partial_{\beta} I$. The action is calculated on shell, since on-shell configuration is the dominant contribution in path integral. On shell integral of the action for the regions outside of the horizon amounts to the volume of the spacetime.  
\begin{align}
    I_{\text{on-shell}} = \frac{d}{8\pi\GN}\int \sqrt{g}\,\, d^{d+1}x 
\end{align}
To calculate vacuum subtracted energy, one should calculate this integral for the AdS-Schwarzschild for region $r_{+}\le r \le r_{\infty}$ and subtract the vacuum contribution by considering the same integral on AdS$_{d+1}$ for $0\le r \le r_{\infty}$.
\begin{align}
I =\lim_{r_{\infty}\to\infty} \frac{d}{8\pi \GN}\left( \text{Vol}_{BH}(r_{\infty})-\text{Vol}_{AdS}(r_{\infty})\right)
\end{align}
Explicit calculation yields,
\begin{align}
    M = \frac{\partial I}{\partial \beta} = \frac{(d-1)\Omega_{d-1}}{16\pi \GN}\left(\frac{r_{+}^d}{L^2}+r_{+}^{d-2}\right)
\end{align}
which correctly reproduces the vacuum subtracted energy of the BTZ black hole $d=2$ ($M_{\text{AdS}_3}=-1/8\GN$). Let us now use this expression to compare the two bounds and find the limit where Araki-Lieb takes over the modular energy bound.  
\begin{align}\label{condition}
    \frac{d-1}{2} \left(\frac{r_{+}}{L}+\frac{L}{r_{+}}\right) \cot{\alpha} \le 1
\end{align}
Therefore when the condition above holds, Bekenstein bound sets a lower bound then the Araki-Lieb. For large black holes \ie\, $r_{+}/ L \gg 1$, the bound puts a more restrictive condition then Araki-Lieb when $\cot{\alpha} \le \frac{2}{d-1} \frac{L}{r_{+}}$. In this case sphere of ignorance stays inside the black hole. Hence the entropy can not only be associated to the sphere $ \mathbb{S}^{d-1}_{R_{\text{scale}}}$ as it is not confined in this region. On the other hand, when $r_{+}/ L < 1$ thermal AdS solution dominates the canonical ensemble. 

The metric of Euclidean thermal AdS solution is identical to empty Euclidean AdS apart from periodicity of time direction $t_{E}\sim t_{E} + \beta$. The difference with Euclidean Schwarzschild is that time circle in this solution does not cap off around the origin, in other words while space $(r, t_E)$ is topologically a disc for BH solution, it is topologically equivalent to a cylinder in thermal AdS. Hence $\beta$ is not fixed by any regularity condition and becomes a free parameter in this solution. The solution is represented as empty AdS therefore holographic entanglement entropy  at the leading order is identical to vacuum entanglement entropy. One can calculate the entropy of the solution via the on-shell action calculated on $\mathbb{R}^{d+1}\times \mathbb{S}^{1}_{\beta}$,  $S = (1-\beta\partial_{\beta}) I = 0$. Thermal entropy on top of the vacuum contribution comes at the order $\mathcal{O}(\GN^{0})$. At the critical temperature where phase transition takes place it suddenly jumps to   $\mathcal{O}(1/\GN)$. Below this phase transition, the thermal entropy can be fully confined inside the $\mathbb{S}^{d-1}_{R_{\text{scale}}}$ as it has been studied for thermal states at the linear level. Boundary computation provides all orders to the perturbative excitation. Hence a valid interpretation of the Bekenstein bound on the entropy attributed to sphere of ignorance takes place below the Hawking-Page transition. This is in agreement with general understanding on the fact that Bekenstein bound is applicable to system having weak self gravitation.

To sum up when condition \eqref{condition} is satisfied the Bekenstein bound proposed in this paper sets a lower bound then the Araki-Lieb bound. One can satisfy this condition in both sides of the critical temperature. When the temperature is above the critical temperature, black hole solutions are dominant in the phase space. In this case, the space of parameters that satisfy \eqref{condition}, have $R_{scale}$ that falls into the black hole, $R_{scale} < L < r_{+}$. On the other regime, below the Hawking-Phase transition, the parameter space satisfy the condition when $r_{+} \leq R_{scale}$. In this case one can confine all the excitation inside the bulk sphere $\mathbb{S}_{R}$. We think thermal AdS regime is more natural for Bekenstein bound interpretation in the bulk since one can push all thermal gas inside the sphere $\mathbb{S}_{R}$. Our understanding also agrees with the general idea that bound is valid for weakly self gravitating systems.

\section{Conclusion and Discussion}
In this paper we have identified the \textit{full modular Hamiltonian} from the bulk point of view. We have studied analogous quantity in the entanglement entropy and shown that it has distinct character for pure and mixed state excitation. In section 2 we show that purity of the state puts strict constraints on the allowed expectation value of the boundary stress energy tensor on the excited state. The bulk interpretation of the full modular Hamiltonian for certain class of excitation have a remarkably simple expression independent of the dimension of the spacetime. The connection between the bulk expression of the full modular hamiltonian ($2 \pi M R$) and the change of area in a certain identification of the manifolds had been used to modify gravity at long distance scales \cite{Verlinde:2016toy}. In that proposal, this expression was the key component in the gravitational side where underlying state follows the area law. Here we have identified the expression as the full modular hamiltonian in the underlying theory. The main conclusions of this paper can be listed in the following way.
\begin{itemize}
    \item \textbf{Bekenstein bound in the bulk:} Using positivity together with the monotonicity of relative entropy,  we have shown that the change of entanglement entropy for complementary states in spherically symmetric excitations are bounded by $2 \pi M R_{\text{scale}}$. The expression is valid perturbatively as well as non perturbatively. In the perturbative regime, full modular hamiltonian can be expressed in the bulk as the integral of the local bulk stress energy tensor using covariant phase space approach. In this case it is clear that bound is saturated if all the excitation is hidden behind the sphere of ignorance defined with respect to the boundary region $A$. We have proposed that difference of the change of the entanglement entropies of complementary regions ($S_{\bar{A}}- S_{A}$) in the boundary theory sets a bound for the entanglement entropy resides in the sphere of ignorance. This entanglement should be seen as the entanglement with respect to the purifying state, which would be zero for pure state excitation, which trivially satisfy the Bekenstein bound proposed here. In conclusion, the Bekenstein bound in the bulk manifest itself as the positivity together with the monotonicity of the relative entropy in the boundary CFT. 

    \item \textbf{An example of UV-IR correspondence:} Bulk interpretation of the full modular Hamiltonian reflects the well known UV-IR correspondence \cite{Susskind:1998dq}. This should be understood as follows; consider the change of full modular Hamiltonian, $\Delta \langle H^{\text{full}}(\alpha) \rangle$ in the boundary CFT as a function of the size of the boundary interval $A$, where $\alpha \in [0, \pi/2]$. In the boundary theory $\alpha \rightarrow 0$ limit identifies the short distance behaviour. On the other hand, as we have seen in the bulk the quantity amounts to $2 \pi M R_{\text{scale}}$, where $R_{\text{scale}}$ denotes the deepest point the bulk that can be probed from the boundary state $A$. The limit $\alpha \rightarrow 0$ corresponds to $R_{\text{scale}} \rightarrow \infty$ from the bulk point of view which yields one realization of the UV-IR correspondence in the AdS/CFT.
    
    \item \textbf{Black hole vs thermal gas limits:} Comparing with the Araki-Lieb bound we have seen that, Bekenstein bound sets a lower bound when the complementary regions are close to each other. In the case of large black holes, when the Bekenstein limit sets the lower bound with respect to Araki-Lieb, the sphere of ignorance to which we have associated the entropy ($S_{\bar{A}}- S_{A}$) corresponds to the regions inside the black hole. In that case holographic bound is already satisfied due to the formation of the black hole. Hence holographic bound sets even a lower bound than the Bekenstein one in these cases. On the other hand, in the thermal gas limit, ( $r_{+}/ L < 1$), below the Hawking-Page transition, one can come up with a window, where Bekenstein bound becomes non-trivial with respect to Araki-Lieb and yet holographic bound is not saturated. This corresponds to limit where self gravitation of the excitation is weak, hence in agreement with the expectations that bound can be derived within the context of QFTs. On the other hand our derivation includes the backreaction on the geometry. We have also shown that in the weak field limit the full modular hamiltonian have well defined bulk expression which further justifies the proposals made in this paper.
    
    \item \textbf{Boundary to bulk map and proof of the proposal:} We find it useful to emphasize that we have not provided the full proof of the derivation of the Bekenstein bound in the bulk via AdS/CFT. The relation between entropy associated to the bulk spheres $\Delta S^{\text{bulk}}_{\mathbb{S}_{R}}$ and boundary entropy difference $\le \Delta S_{\bar{A}-A}$ is conjectured for a spherically symmetric state \eqref{bulkboundaryentropyconjecture}. We have proved this conjecture in the perturbative limit. Under such an assumption we show that Bekenstein bound in the bulk manifest itself as the information inequalities (positivity + monotonicity) in the underlying theory. It would be remarkable to find the exact map between entropy of the bulk regions and boundary regions to drop the assumption of spherical symmetry. That would also let us test the volume law conjecture in spacetime.

\end{itemize}

\section*{Acknowledgement}

I would like to thank Nava Gaddam, and Erik Verlinde for interesting discussions and Krzysztof Sadowski for comments on the manuscript. This work is supported by the Spinoza grant funded by the Dutch science organization NWO.

\newpage
\renewcommand{\leftmark}{\MakeUppercase{Bibliography}}
\phantomsection
\bibliographystyle{JHEP}
\bibliography{bekenstein_bound_and_ads_cft}

\providecommand{\href}[2]{#2}\begingroup\raggedright\begin{thebibliography}{10}

\bibitem{Bekenstein:1980jp}
J.~D. Bekenstein, \emph{{A Universal Upper Bound on the Entropy to Energy Ratio
  for Bounded Systems}},
  \href{https://doi.org/10.1103/PhysRevD.23.287}{\emph{Phys. Rev.} {\bfseries
  D23} (1981) 287}.

\bibitem{Bekenstein:1974ax}
J.~D. Bekenstein, \emph{{Generalized second law of thermodynamics in black hole
  physics}}, \href{https://doi.org/10.1103/PhysRevD.9.3292}{\emph{Phys. Rev.}
  {\bfseries D9} (1974) 3292}.

\bibitem{Casini:2008cr}
H.~Casini, \emph{{Relative entropy and the Bekenstein bound}},
  \href{https://doi.org/10.1088/0264-9381/25/20/205021}{\emph{Class. Quant.
  Grav.} {\bfseries 25} (2008) 205021}
  [\href{https://arxiv.org/abs/0804.2182}{{\ttfamily 0804.2182}}].

\bibitem{Maldacena:1997re}
J.~M. Maldacena, \emph{{The Large N limit of superconformal field theories and
  supergravity}}, \href{https://doi.org/10.1023/A:1026654312961,
  10.4310/ATMP.1998.v2.n2.a1}{\emph{Int. J. Theor. Phys.} {\bfseries 38} (1999)
  1113} [\href{https://arxiv.org/abs/hep-th/9711200}{{\ttfamily
  hep-th/9711200}}].

\bibitem{Witten:1998qj}
E.~Witten, \emph{{Anti-de Sitter space and holography}},
  \href{https://doi.org/10.4310/ATMP.1998.v2.n2.a2}{\emph{Adv. Theor. Math.
  Phys.} {\bfseries 2} (1998) 253}
  [\href{https://arxiv.org/abs/hep-th/9802150}{{\ttfamily hep-th/9802150}}].

\bibitem{Aharony:1999ti}
O.~Aharony, S.~S. Gubser, J.~M. Maldacena, H.~Ooguri and Y.~Oz, \emph{{Large N
  field theories, string theory and gravity}},
  \href{https://doi.org/10.1016/S0370-1573(99)00083-6}{\emph{Phys. Rept.}
  {\bfseries 323} (2000) 183}
  [\href{https://arxiv.org/abs/hep-th/9905111}{{\ttfamily hep-th/9905111}}].

\bibitem{Bisognano:1976za}
J.~J. Bisognano and E.~H. Wichmann, \emph{{On the Duality Condition for Quantum
  Fields}}, \href{https://doi.org/10.1063/1.522898}{\emph{J. Math. Phys.}
  {\bfseries 17} (1976) 303}.

\bibitem{Ryu:2006bv}
S.~Ryu and T.~Takayanagi, \emph{{Holographic derivation of entanglement entropy
  from AdS/CFT}},
  \href{https://doi.org/10.1103/PhysRevLett.96.181602}{\emph{Phys. Rev. Lett.}
  {\bfseries 96} (2006) 181602}
  [\href{https://arxiv.org/abs/hep-th/0603001}{{\ttfamily hep-th/0603001}}].

\bibitem{Ryu:2006ef}
S.~Ryu and T.~Takayanagi, \emph{{Aspects of Holographic Entanglement Entropy}},
  \href{https://doi.org/10.1088/1126-6708/2006/08/045}{\emph{JHEP} {\bfseries
  08} (2006) 045} [\href{https://arxiv.org/abs/hep-th/0605073}{{\ttfamily
  hep-th/0605073}}].

\bibitem{Lewkowycz:2013nqa}
A.~Lewkowycz and J.~Maldacena, \emph{{Generalized gravitational entropy}},
  \href{https://doi.org/10.1007/JHEP08(2013)090}{\emph{JHEP} {\bfseries 08}
  (2013) 090} [\href{https://arxiv.org/abs/1304.4926}{{\ttfamily 1304.4926}}].

\bibitem{Iyer:1994ys}
V.~Iyer and R.~M. Wald, \emph{{Some properties of Noether charge and a proposal
  for dynamical black hole entropy}},
  \href{https://doi.org/10.1103/PhysRevD.50.846}{\emph{Phys. Rev.} {\bfseries
  D50} (1994) 846} [\href{https://arxiv.org/abs/gr-qc/9403028}{{\ttfamily
  gr-qc/9403028}}].

\bibitem{Verlinde:2016toy}
E.~P. Verlinde, \emph{{Emergent Gravity and the Dark Universe}},
  \href{https://arxiv.org/abs/1611.02269}{{\ttfamily 1611.02269}}.

\bibitem{deBoer:2015kda}
J.~de~Boer, M.~P. Heller, R.~C. Myers and Y.~Neiman, \emph{{Holographic de
  Sitter Geometry from Entanglement in Conformal Field Theory}},
  \href{https://doi.org/10.1103/PhysRevLett.116.061602}{\emph{Phys. Rev. Lett.}
  {\bfseries 116} (2016) 061602}
  [\href{https://arxiv.org/abs/1509.00113}{{\ttfamily 1509.00113}}].

\bibitem{Faulkner:2013ana}
T.~Faulkner, A.~Lewkowycz and J.~Maldacena, \emph{{Quantum corrections to
  holographic entanglement entropy}},
  \href{https://doi.org/10.1007/JHEP11(2013)074}{\emph{JHEP} {\bfseries 11}
  (2013) 074} [\href{https://arxiv.org/abs/1307.2892}{{\ttfamily 1307.2892}}].

\bibitem{Maldacena:2013xja}
J.~Maldacena and L.~Susskind, \emph{{Cool horizons for entangled black holes}},
  \href{https://doi.org/10.1002/prop.201300020}{\emph{Fortsch. Phys.}
  {\bfseries 61} (2013) 781} [\href{https://arxiv.org/abs/1306.0533}{{\ttfamily
  1306.0533}}].

\bibitem{Blanco:2017akw}
D.~Blanco, H.~Casini, M.~Leston and F.~Rosso, \emph{{Modular energy
  inequalities from relative entropy}},
  \href{https://doi.org/10.1007/JHEP01(2018)154}{\emph{JHEP} {\bfseries 01}
  (2018) 154} [\href{https://arxiv.org/abs/1711.04816}{{\ttfamily
  1711.04816}}].

\bibitem{Herzog:2014fra}
C.~P. Herzog, \emph{{Universal Thermal Corrections to Entanglement Entropy for
  Conformal Field Theories on Spheres}},
  \href{https://doi.org/10.1007/JHEP10(2014)028}{\emph{JHEP} {\bfseries 10}
  (2014) 28} [\href{https://arxiv.org/abs/1407.1358}{{\ttfamily 1407.1358}}].

\bibitem{Cardy:2016fqc}
J.~Cardy and E.~Tonni, \emph{{Entanglement hamiltonians in two-dimensional
  conformal field theory}},
  \href{https://doi.org/10.1088/1742-5468/2016/12/123103}{\emph{J. Stat. Mech.}
  {\bfseries 1612} (2016) 123103}
  [\href{https://arxiv.org/abs/1608.01283}{{\ttfamily 1608.01283}}].

\bibitem{Deser:1983tn}
S.~Deser, R.~Jackiw and G.~'t~Hooft, \emph{{Three-Dimensional Einstein Gravity:
  Dynamics of Flat Space}},
  \href{https://doi.org/10.1016/0003-4916(84)90085-X}{\emph{Annals Phys.}
  {\bfseries 152} (1984) 220}.

\bibitem{Casini:2011kv}
H.~Casini, M.~Huerta and R.~C. Myers, \emph{{Towards a derivation of
  holographic entanglement entropy}},
  \href{https://doi.org/10.1007/JHEP05(2011)036}{\emph{JHEP} {\bfseries 05}
  (2011) 036} [\href{https://arxiv.org/abs/1102.0440}{{\ttfamily 1102.0440}}].

\bibitem{Cardy:2014jwa}
J.~Cardy and C.~P. Herzog, \emph{{Universal Thermal Corrections to Single
  Interval Entanglement Entropy for Two Dimensional Conformal Field Theories}},
  \href{https://doi.org/10.1103/PhysRevLett.112.171603}{\emph{Phys. Rev. Lett.}
  {\bfseries 112} (2014) 171603}
  [\href{https://arxiv.org/abs/1403.0578}{{\ttfamily 1403.0578}}].

\bibitem{Popescu2005}
S.~Popescu, A.~J. Short and A.~Winter, \emph{{Entanglement and the foundations
  of statistical mechanics}},
  \href{https://doi.org/doi:10.1038/nphys444}{\emph{Nature Physics} {\bfseries
  2} (2006) 754}.

\bibitem{Faulkner:2013ica}
T.~Faulkner, M.~Guica, T.~Hartman, R.~C. Myers and M.~Van~Raamsdonk,
  \emph{{Gravitation from Entanglement in Holographic CFTs}},
  \href{https://doi.org/10.1007/JHEP03(2014)051}{\emph{JHEP} {\bfseries 03}
  (2014) 051} [\href{https://arxiv.org/abs/1312.7856}{{\ttfamily 1312.7856}}].

\bibitem{Swingle:2014uza}
B.~Swingle and M.~Van~Raamsdonk, \emph{{Universality of Gravity from
  Entanglement}},  \href{https://arxiv.org/abs/1405.2933}{{\ttfamily
  1405.2933}}.

\bibitem{Lin:2014hva}
J.~Lin, M.~Marcolli, H.~Ooguri and B.~Stoica, \emph{{Locality of Gravitational
  Systems from Entanglement of Conformal Field Theories}},
  \href{https://doi.org/10.1103/PhysRevLett.114.221601}{\emph{Phys. Rev. Lett.}
  {\bfseries 114} (2015) 221601}
  [\href{https://arxiv.org/abs/1412.1879}{{\ttfamily 1412.1879}}].

\bibitem{Dong:2016eik}
X.~Dong, D.~Harlow and A.~C. Wall, \emph{{Reconstruction of Bulk Operators
  within the Entanglement Wedge in Gauge-Gravity Duality}},
  \href{https://doi.org/10.1103/PhysRevLett.117.021601}{\emph{Phys. Rev. Lett.}
  {\bfseries 117} (2016) 021601}
  [\href{https://arxiv.org/abs/1601.05416}{{\ttfamily 1601.05416}}].

\bibitem{Faulkner:2017vdd}
T.~Faulkner and A.~Lewkowycz, \emph{{Bulk locality from modular flow}},
  \href{https://doi.org/10.1007/JHEP07(2017)151}{\emph{JHEP} {\bfseries 07}
  (2017) 151} [\href{https://arxiv.org/abs/1704.05464}{{\ttfamily
  1704.05464}}].

\bibitem{Bekenstein:2003dt}
J.~D. Bekenstein, \emph{{Black holes and information theory}},
  \href{https://doi.org/10.1080/00107510310001632523}{\emph{Contemp. Phys.}
  {\bfseries 45} (2003) 31}
  [\href{https://arxiv.org/abs/quant-ph/0311049}{{\ttfamily
  quant-ph/0311049}}].

\bibitem{Leichenauer:2015nxa}
S.~Leichenauer, \emph{{Thermal Corrections to Entanglement Entropy from
  Holography}}, \href{https://doi.org/10.1007/JHEP09(2015)014}{\emph{JHEP}
  {\bfseries 09} (2015) 014}
  [\href{https://arxiv.org/abs/1502.07348}{{\ttfamily 1502.07348}}].

\bibitem{Freedman:2016zud}
M.~Freedman and M.~Headrick, \emph{{Bit threads and holographic entanglement}},
  \href{https://doi.org/10.1007/s00220-016-2796-3}{\emph{Commun. Math. Phys.}
  {\bfseries 352} (2017) 407}
  [\href{https://arxiv.org/abs/1604.00354}{{\ttfamily 1604.00354}}].

\bibitem{Hubeny:2013gta}
V.~E. Hubeny, H.~Maxfield, M.~Rangamani and E.~Tonni, \emph{{Holographic
  entanglement plateaux}},
  \href{https://doi.org/10.1007/JHEP08(2013)092}{\emph{JHEP} {\bfseries 08}
  (2013) 092} [\href{https://arxiv.org/abs/1306.4004}{{\ttfamily 1306.4004}}].

\bibitem{Araki1970aheh}
H.~Araki and E.~H. Lieb, \emph{{Entropy Inequalities}},
  \href{https://doi.org/10.1007/BF01646092}{\emph{Commun.Math. Phys.}
  {\bfseries 18} (1970) 160}.

\bibitem{Hawking:1982dh}
S.~W. Hawking and D.~N. Page, \emph{{Thermodynamics of Black Holes in anti-De
  Sitter Space}}, \href{https://doi.org/10.1007/BF01208266}{\emph{Commun. Math.
  Phys.} {\bfseries 87} (1983) 577}.

\bibitem{Witten:1998zw}
E.~Witten, \emph{{Anti-de Sitter space, thermal phase transition, and
  confinement in gauge theories}},
  \href{https://doi.org/10.4310/ATMP.1998.v2.n3.a3}{\emph{Adv. Theor. Math.
  Phys.} {\bfseries 2} (1998) 505}
  [\href{https://arxiv.org/abs/hep-th/9803131}{{\ttfamily hep-th/9803131}}].

\bibitem{Susskind:1998dq}
L.~Susskind and E.~Witten, \emph{{The Holographic bound in anti-de Sitter
  space}},  \href{https://arxiv.org/abs/hep-th/9805114}{{\ttfamily
  hep-th/9805114}}.

\end{thebibliography}\endgroup

\end{document}